\documentclass[3p]{elsarticle}
\usepackage[utf8]{inputenc}
\usepackage{amsmath, graphicx, verbatim, dsfont, amsfonts, color}
\usepackage{braket}
\usepackage{lineno,hyperref}
\usepackage[us]{datetime}
\modulolinenumbers[5]
\begin{document}

\title{2D electron gas in chalcogenide multilayers}

\author{A.~Kazakov}
\author{T.~Wojtowicz}
\address{International Research Centre MagTop, Institute of Physics, Polish Academy of Sciences, Aleja Lotnikow 32/46, PL-02668 Warsaw, Poland}

\begin{abstract}
Semiconductor interfaces, such as these existing in multilayer structures (e.g., quantum wells (QWs)), are interesting because of their ability to form 2D electron gases (2DEGs), in which charge carriers behave completely differently than they do in the bulk. As an example, in the presence of a strong magnetic field, the Landau quantization of electronic levels in the 2DEG results in the quantum Hall effect (QHE), in which Hall conductance is quantized. This chapter is devoted to the properties of such 2DEGs in multilayer structures made of compound semiconductors belonging to the class of Se- and Te-based chalcogenides. In particular, we will also discuss the interesting question of how the QHE phenomenon is affected by the giant Zeeman splitting characteristic of II-VI-based diluted magnetic semiconductors (DMSs), especially when the Zeeman splitting and Landau splitting become comparable. We will also shortly discuss novel topological phases in chalcogenide multilayers.
\end{abstract}

\maketitle


\section{Introduction}

In condensed matter physics, a two-dimensional electron gas (2DEG) is a model system to study electron motion in reduced dimensions. Electrons are free to move in two dimensions but are strictly confined in the third. Such confinement leads to quantized energy levels in the third dimension. These conditions are usually achieved in a quantum well (QW) or transistor-like semiconductor heterostructures and recently in atomically-thin 2D materials. At low temperatures, the mobility of a 2DEG can reach extremely high values, up to several $10^7~cm^2/Vs$ for GaAs based QWs \cite{Umansky2009,Manfra2014}. Such high-quality QWs opened access to new physics, such as integer \cite{Klitzing1980} and fractional \cite{Tsui1982} quantum Hall effects, fractional statistics \cite{Arovas1984}, and many other phenomena. In QWs made of chalcogenides, carrier mobility is generally lower than found in III-V semiconductor QWs. However, the advantage of chalcogenide QWs is that they can be doped with Mn without affecting the carrier concentration and without strongly reducing carrier mobility. Manganese is a neutral impurity in II-VI semiconductors but is an acceptor in III-V semiconductors. The exchange interaction between conduction electrons and magnetic ions give rise to a giant spin splitting. Thus, chalcogenide QWs made of diluted magnetic semiconductor (DMS) materials offer the possibility to study highly spin-polarized low-density 2DEGs \cite{Smorchkova1997,Teran2002,Jaroszynski2002}. In addition, QWs based on narrow-gap semiconductors, such as HgTe, offer 2DEG with a giant spin-orbit splitting \cite{Gui2004b} and the possibility to study a 2D system, where both electrons and holes coexist \cite{Kvon2008}.

2DEGs in DMS-based QWs have been the subject of intensive research in terms of possible applications in spintronic devices \cite{Gruber2001,Slobodskyy2003,Betthausen2012}. Different devices implementing giant spin splitting have been proposed, e.g., spin aligners and injectors. Among the proposed spin injectors is a device based on a resonant tunneling diode (RTD) \cite{Gruber2001,Slobodskyy2003}, consisting of a (Zn,Mn)Se DMS QW sandwiched between thin, non-magnetic $\text{Zn}_{1-x}\text{Be}_x\text{Se}$ barriers, which were separating QW from ZnSe epilayers. In such a device one can selectively bring the spin-up or spin-down states into resonance thus increasing the transmission probability of electrons with selected spin alignment.

In recent years there is renewed interest in 2DEG in chalcogenides due to the prediction of various topological phases in such compounds. The quantum spin Hall effect (QSHE), which is a hallmark of topological insulator (TI) phase, was realized in HgTe QWs \cite{Koenig2007}. Ferromagnetism introduced into a suitable TI can break time-reversal symmetry and lead to the quantum anomalous Hall effect (QAHE). It also was experimentally realized in 2D systems based on chalcogenides, namely in $\text{Cr}_{0.15}(\text{Bi}_{0.1}\text{Sb}_{0.9})_{1.85}\text{Te}_3$ \cite{Chang2013} and in $\text{Hg}_{1-x}\text{Mn}_{x}\text{Te}$ QWs \cite{Budewitz2017}. One of the main focus in the research of topological materials is the realization of so-called topological superconductor (TSC) phase, with boundary states obeying non-Abelian statistics. Though at the moment there is no readily available material which is in the TSC phase (except the debated occurrence of the TSC phase in $\text{Sr}_2\text{Ru}\text{O}_4$ \cite{Kallin2016,Sato2017}), it can be realized in hybrid semiconductor-superconductor systems. Many different proposals have been developed in recent years, several among them are based on DMS QWs \cite{Fatin2016,Simion2018}.

Most of the results obtained on 2D and bulk DMS systems in the 1990s and 2000s can be found in extensive reviews \cite{Dobrowolski2003,Gaj2010}, while transport studies in HgTe and PbTe QWs have been reported in Refs. \cite{Kvon2009,Kvon2017} and \cite{Grabecki2007}, respectively. In this review, we will focus mainly on experimental results and proposals published in recent years. In the first part, we will concentrate on the consequence of giant spin splitting for magnetotransport. The second part of the current review is devoted to topological phases found in chalcogenide QWs. In particular, we will review several concepts concerning the creation of Majorana states in CdTe QWs and briefly mention QSHE and QAHE in HgTe QWs. We will not discuss here 2D TI phases found in $(\text{Bi},\text{Sb})_2\text{Te}_3$ and $(\text{Bi},\text{Sb})_2\text{Se}_3$ thin films, where the top and bottom surface states hybridize.

\section{2DEG in magnetically doped QWs}

\subsection{2DEG in low-dimensional heterostructures}

Most of the chalcogenide heterostructures hosting 2DEG are based on the II-VI semiconductors, such as ZnSe(ZnTe), CdTe, and HgTe. All of them have a zinc-blende crystal structure with direct band gap at the $\Gamma$ point. ZnSe and ZnTe have the low temperature highest band gap values of 2.82 and 2.40~eV respectively. The band gap of CdTe is equal to 1.6~eV, while HgTe has the so-called inverted band structure with the band gap of -0.3~eV. It is worth to mention that by varying the composition of $\text{Hg}_{1-x}\text{Cd}_{x}\text{Te}$, the band gap also varies from a positive to a negative value through a zero gap for a particular composition. It was found in Ref. \cite{Galazka1967} that such zero gap energy spectrum in $\text{Hg}_{0.9}\text{Cd}_{0.1}\text{Te}$ has linear dispersion $\epsilon\propto{\bf{k}}$, i.e., Dirac cone. Refs. \cite{Keldysh1964,Wolff1964,Zawadzki1966,Aronov1967} were the first, which recognized the formal analogy between the dispersion relation $\epsilon({\bf{k}})$ of relativistic electrons in vacuum and electrons in semiconductors, where the parameter corresponding to the light velocity is proportional to the band gap. After years of development \cite{Zawadzki2017}, this analogy lead the discovery and understanding of graphene and topological insulators (TIs). The concept of TIs is discussed at the end of the current chapter (sec. \ref{sQSHE_HgTe}).

Other material systems which we will only briefly mention are IV-VI semiconductors, such as (Pb,Sn)Te and (Pb,Sn)Se. (Pb,Sn)Te and (Pb,Sn)Se crystallize in the cubic rock-salt structure, provided that Sn content is low enough. Their multivalley band structure consists of four elongated ellipsoidal valleys at the L-points of the first Brillouin zone, one for each equivalent [111] direction. The growth of QWs is usually done on (111) $\text{BaF}_2$ along the [111] crystallographic axis, which results in lifting the fourfold L-valley degeneracy. Therefore, the ground-state of the 2D subband originates from a single valley with the long axis oriented along the [111] growth direction (longitudinal valley), and the lowest states formed by another three valleys obliquely oriented to the growth direction at an angle of $70.53^\circ$ (oblique valleys) has higher energy than the longitudinal valley first excited state \cite{Yuan1994,Abramof2001}. It is worth to mention that low temperature band gap value in PbTe is 0.18~eV and SnTe has an inverted band structure with -0.3~eV band gap value \cite{Dimmock1966}. Variation of Tin content in $\text{Pb}_{1-x}\text{Sn}_{x}\text{Te}$ ($\text{Pb}_{1-x}\text{Sn}_{x}\text{Se}$) alloy result in similar physics as for $\text{Hg}_{1-x}\text{Cd}_{x}\text{Te}$ alloys, i.e., linear dispersion law.

2DEG in chalcogenide materials is usually achieved in heterostructures containing a QW. These heterostructures are made of semiconductors with unequal band gaps, so the resulting band structure can be adjusted to one's need by an engineered layering of different materials \cite{Henini2013}. Such layered structures are usually fabricated with the use of the molecular beam epitaxy (MBE) technique, which allows growing thin films with atomic precision. To restrict the motion of carriers to a 2D plane a confining potential is needed. Confinement is achieved either in type I heterojunctions together with doping of the wider band gap semiconductor, or in QW structures. In a QW structure, a lower band gap semiconductor layer is sandwiched between wider band gap semiconductor layers. A standard way to separate free charge carriers and charged impurities is the modulation-doping scheme \cite{Dingle1978}. In such a scheme, dopants are spatially separated from the confining potential well. Thus, carrier scattering at ionized impurities (donors and acceptors) is greatly reduced, which results in higher low-temperature carrier mobility --- a conventional measure of 2DEG quality.

In most of the works devoted to QWs based on wide band-gap II-VI semiconductors, the QW was grown along the [001] direction on a GaAs substrate. Though there have been studies of QWs grown along other crystallographic directions, they are rather scarce. The big progress in realizing 2D systems based on II-VI materials was achieved after the introduction of halogen dopants. The main advantage of halogen doping over previously more traditional indium doping is its smaller tendency to produce DX-like centers with deep in-gap states \cite{Wasik1999} and higher doping efficiency \cite{Scholl1993}. Implementation of halogen doping allowed to create 2DEGs in non-magnetic QWs CdTe/CdMgTe \cite{Scholl1995,Karczewski1998}, ZnTe/CdSe \cite{Smorchkova1998} as well as in their DMS analogues: CdTe/CdMnTe \cite{Scholl1993}, ZnSe/(Zn,Cd,Mn)Se \cite{Smorchkova1996,Smorchkova1997}, CdMnTe/CdMgTe \cite{Wojtowicz1998}, ZnTe/CdMnSe \cite{Knobel1999}, etc. In the case of DMS QWs, the magnetic ions (Mn ions are usually used for this purpose) can be placed either into the barrier materials or into the QWs themselves. In both cases, the electron wavefunction in the QW overlaps with that of the magnetic ion giving rise to s-d exchange interaction, which modifies QW energy levels, especially in the presence of an external magnetic field. There are two approaches to place magnetic dopants into QW. The first is the so-called "disordered alloy", in which the magnetic impurities are randomly distributed over the whole QW width. In the other approach \cite{Crooker1995,Wojtowicz1995}, called "digital alloy", nonmagnetic layers (e.g., CdTe or ZnCdSe) are separated by atomically thin magnetic layers (e.g., CdMnTe or MnSe). Such a technique allows achieving higher local concentrations of magnetic ions without the transition to antiferromagnetic or spin glass phases \cite{Smorchkova1997}. It was also observed that in various versions of "digital alloy" and "disordered alloy" doping schemes, for the same number of incorporated paramagnetic spins (resulting in the same average concentration of Mn in the QW) the spin splitting of excitonic states is the same while the magnetization dynamics is different \cite{Kneip2006b}. The mobilities of the 2DEGs in digital alloy QWs were high enough to allow observation of Shubnikov-de Haas (SdH) oscillations in magnetoresistance (MR) and the quantum Hall effect (QHE).

Here it is worth to note some peculiarities of the band alignment in HgTe/HgCdTe QWs. Calculations performed with the eight-band ${\bf k}\cdot {\bf p}$ method \cite{Novik2005,Becker2007} have shown that the positions of E1 and H1 bands --- E and H label the electron and heavy hole subbands --- strongly depend on the QW thickness, $d$. Above a critical thickness, $d_c\simeq6.3$~nm H1 lays above E1 so the band structure becomes inverted. In the case of $d\simeq d_c$ linear dispersion relation, i.e., Dirac cone is observed, thus making HgTe QW with critical thickness a single valley Dirac semimetal \cite{Buettner2011}. HgTe QWs are usually grown in [001] direction with iodine or indium doping, however, there are several works devoted to (013) QWs \cite{Kvon2008,Kvon2009}. Due to the difference between HgTe and CdTe lattice constants, HgTe QWs are strained. It was found that in wide (013) HgTe QWs with $d\sim20$~nm \cite{Kvon2011} this strain leads to overlap between H1 and H2 bands with extrema points at different {\bf k}, thus constituting a 2D electron-hole system. 

The most investigated IV-VI semiconductor QWs are PbTe/PbEuTe heterostructures \cite{Olver1994,Chitta2005}. They are usually grown on (111) $\text{BaF}_2$ substrate and Bi is used as an {\it n}-type dopant \cite{Rappl1998,Springholz2013}. Without intentional doping IV-VI semiconductors are typically {\it p}-type due to non-stoichiometry; cation vacancies are a natural source of holes \cite{Parada1969}. PbTe has been chosen as a QW material since it possesses the highest mobility among lead-tin selenides and tellurides \cite{Khokhlov2002}. In addition, PbTe has a very large dielectric constant ($\epsilon_{\text{PbTe}}=1300-3000$ at helium temperatures \cite{Nimtz1983}), hence the Coulomb potential from defects and impurities is efficiently screened. For this reason doping within QW hardly affects the mobility, thus making remote doping unnecessary. The high mobility makes PbTe QWs perfect to study quasiballistic transport phenomena \cite{Grabecki2002,Grabecki2005,Grabecki2007}. Already the first PbTe/PbEuTe QWs were of good enough quality to allow for the observation of SdH oscillations and signatures of QHE \cite{Olver1994}. Nevertheless, studies of QWs based on these materials remain rather scarce.

It is known that doping of QWs induces a certain disorder which affects the transport properties of 2DEG. This disorder is partially responsible for the persistent photoconductivity (PPC) effect which is observed after illumination of a semiconductor at low temperature. QWs based on II-VI semiconductors exhibit similar effects \cite{Ray1999,Kazakov2017}. In the work of Ray {\it et. al.} \cite{Ray1999} PPC in magnetically doped ZnSe QW was interpreted as due to the formation of DX centers at Mn sites. However, it is known that {\it n}-type dopants also create DX centers in CdTe, CdMnTe, and CdMgTe \cite{Iseler1972,Wojtowicz1993,Wojtowicz1994,Semaltianos1993,Wasik1999,Wasik1996}. Temperature dependent PPC was also observed in PbTe/PbEuTe QWs with contributions from both barriers and QW \cite{Pena2017}. At high temperatures transport via barriers was dominating, while in the low-temperature range conduction proceeded within the QW.

\subsection{Spin interactions in chalcogenide DMS QWs}

As already mentioned, the advantage of II-VI chalcogenides over III-V semiconductors is the fact that Manganese (Mn) form a neutral impurity in II-VI compounds, though providing local magnetic moments. Mn has the highest magnetic moment $S=5/2$ among transition metals because of its half-filled {\it d}-shell. The direct exchange interaction between neighboring Mn ions is antiferromagnetic, but here we discuss only systems with low Mn content and no magnetic ordering. The giant spin splitting of a 2DEG in the magnetic field stems from the exchange interaction between {\it d}-electrons, which are localized on Mn ions, and conduction {\it s}- and {\it p}-electrons. Exchange contribution to the spin-splitting energy depends on magnetization $M$ of the Mn-subsystem in the applied magnetic field $B$ \cite{Furdyna1988}:
\begin{equation}
M = x_{\text{eff}}N_0g\mu_BS\mathfrak{B}_s\left(\frac{g\mu_BSB}{k_b(T+T_{AF})}\right),
\label{eqMn_mag}
\end{equation}
Here $g=2$ is the free electron $g$-factor; $\mu_B$ is the Bohr magneton; $N_0$ is the total number of cations per unit volume; $\mathfrak{B}_s\left(B,T\right)$ stands for the Brillouin function; $x_{\text{eff}}$ is the effective Mn concentration which is lower than the actual concentration $x_{\text{Mn}}$ due to antiferromagnetic correlations between Mn magnetic moments at higher doping levels \cite{Ferrand2001}; $T_{AF}>0$ accounts for the effect of long-range antiferromagnetic interactions between Mn pairs \cite{Gaj1979}.

In reduced dimensions, under energy confinement conditions, the exchange interaction constant may differ from its bulk value. Indeed, modification of exchange parameters with QW thickness was observed in $\text{Cd}_{1-x}\text{Mn}_{x}\text{Te}/\text{Cd}_{1-x-y}\text{Mn}_{x}\text{Mg}_{y}\text{Te}$ heterostructures with $x_{\text{Mn}}\approx0.04-0.11$ \cite{Mackh1996}. These results agree with those of Ref. \cite{Merkulov1999}, where an explanation of such an effect was suggested. The exchange interaction between conduction {\it s}-electrons from the center of the Brillouin zone and localized {\it d}-electrons has potential character and is positive. For electrons away from the $\Gamma$ point symmetry is no longer purely {\it s}-wave, because {\it p}-like states in the valence band admix to Bloch functions of the conduction band. This admixture introduces kinetic exchange correction to the exchange constant, and the magnitude of this correction depends on the degree of admixture. In the case of narrow QWs, kinetic exchange correction can even lead to a sign reversal of the exchange parameter. However, in recently measured $\text{Cd}_{1-x}\text{Mn}_{x}\text{Te}/\text{Cd}_{0.7}\text{Mg}_{0.3}\text{Te}$ QWs with much lower Mn content ($x\sim0.00146$ and $\sim0.00027$) no change in exchange constants was observed \cite{Rice2012}. Magnetooptical investigations revealed also a pronounced anisotropy of the in-plane Zeeman effect \cite{Peyla1993,Kuhn-Heinrich1994} due to the valence band mixing. The anisotropy of the transverse hole effective $g$-factor $g^*_{\perp}$ in CdTe/CdMnTe QWs could be so large that it can even change the sign for different directiosn of the in-plane magnetic field \cite{Kusrayev1999}.

Besides the exchange interactions, the carrier spins in the QWs experience also spin-orbit (SO) coupling. SO splitting of band states occurs in solids lacking a center of inversion, such as crystals with a zinc-blende structure (Dresselhaus term \cite{Dresselhaus1955}), as well as systems with structural inversion asymmetry (Rashba term \cite{Bychkov1984}), e.g., asymmetric QWs. SO coupling acts on a carrier spin as an effective magnetic field, which strength and direction depend on electron motion.

Dynamics of carrier spins and localized magnetic moments are usually considered together since they are coupled to each other either directly, through exchange scattering \cite{Crooker1995,Baumberg1994,Kneip2006}, or indirectly, through the lattice (phonons) \cite{Kneip2006a}. The first process is quite fast with characteristic times of the order of $10^{-12}-10^{-11}$~s for electrons \cite{Crooker1995,Baumberg1994} and one order of magnitude slower for holes \cite{Baumberg1994,Akimov2006}, as established by time-resolved Faraday rotation measurements. Spin-lattice relaxation times were found to depend strongly on Mn content, ranging from $10^{-3}$~s (for $x_{\text{Mn}}\approx0.4\%$) to $10^{-8}$~s (for $x_{\text{Mn}}\approx10\%$) \cite{Kneip2006a}.

Spin excitations in CdMnTe QWs were probed by the Raman scattering technique and their collective character was established \cite{Teran2003}. A very large Knight shift was observed in the 10-nm CdMnTe QWs ($x_{\text{Mn}}\sim0.2\%$ and $\sim0.3\%$) for a specific magnetic field strength at which the energies of free carrier and Mn spin excitations were almost equal. This ferromagnetic coupling between Mn spins and conduction electrons was theoretically described in Ref. \cite{Koenig2003}. The mean-field description of spin excitations developed in this work agreed quantitatively with the experimental data of Ref. \cite{Teran2003}. It was noted, however, that the long-range magnetic order in (Cd,Mn)Te QWs can be stabilized only by the spin-orbit coupling.

Resonant Raman scattering investigations also proved to be useful in a study of both collective and single-particle spin-flip excitations \cite{Jusserand2003,Perez2006}. Careful measurements of angle-resolved magneto-Raman scattering on spin-flip waves in CdMnTe QWs \cite{Gomez2010} allowed clarification of the dispersion law of the damping rate $\eta$, which was found to be $\eta=\eta_0+\eta_2{\bf q}^2$ \cite{Hankiewicz2008}, where {\bf q} is the spin-wave in-plane momentum. The quadratic in {\bf q} enhancement of the damping rate is described by a single-particle dynamics and occurs because of disorder and spin-Coulomb drag. Moreover, it was found that SO coupling affects spin waves in DMS QWs reconstructing them to chiral spin waves, which are invariant under inversion of both magnetic field and wavevector \cite{DAmico2019}. Observation of chiral spin waves in DMS QWs is possible when Rashba SO coupling is of the order of spin-splitting energy. This condition may be fulfilled by adjusting the QW width, carrier density, and $x_{\text{eff}}$. Indeed, under this condition, chiral spin waves were observed in CdMnTe QWs \cite{Perez2016}. It was found that the dispersions of energy and damping rate are both shifted by a wave vector {\bf q}$_0$, which depends on the relative strength of the SO field. Thus, a change of a SO coupling strength (e.g., through gating or PPC effect) can alter the group velocity of spin waves in a system. Control over the propagation direction of spin waves can find its application in magnonic devices. Spin waves observed in CdMnTe QWs were also found to be a perspective source of spin-based THz radiation \cite{Rungsawang2013}.

In Refs. \cite{Korenev2015,Akimov2017} it was found that interplay between localized moments, phonons, and carriers can even lead to long-range, ferromagnet-induced proximity effect in a ferromagnet-semiconductor hybrid system. In the studied system, a cobalt layer was deposited on the cap layer of CdTe/CdMgTe QW which did not contain any magnetic doping and spin polarization of the acceptor bound hole was detected through photoluminescence spectra. The thickness of the cap layer was varied up to 30~nm which should rule out any proximity effect due to the overlap of the carrier wavefunction with the magnetic layer. It was suggested that elliptically polarized phonons are responsible for angular momentum transfer between the Co layer and the semiconductor leading to polarization of hole spins due to strong SO interaction in the valence band. Polarized phonons couple to transitions between split heavy-hole and light-hole states. This effect is absent in the case of electron gas because the SO interaction is much weaker in the conduction band. The long-range p-d exchange constant was directly measured by spin-flip Raman scattering \cite{Akimov2017} and its value was found to be $50-100~\mu eV$. Application of electric field across such a ferromagnet-semiconductor hybrid structure changes the splitting of the heavy-hole and light-hole states thus affecting phonon coupling. It was shown \cite{Korenev2018} that electric field strengths of the order of $10^4~V/cm$ are sufficient to bring heavy-to-light hole transition out of resonance with the magnon-phonon resonance of the ferromagnet. Therefore, low voltage electric control of the conceptually new long-range exchange coupling mediated by elliptically polarized phonons is possible. 

\subsection{Magnetotransport in chalcogenide QWs}

\subsubsection{Low-field magnetotransport in DMS QWs}

The first grown DMS QWs had usually low mobilities \cite{Scholl1995,Smorchkova1997}, which was caused mainly by poorly developed growth technology and in part by disorder introduced by magnetic ions. All of these QWs had similar features in magnetoresistance (MR) \cite{Scholl1993,Smorchkova1996,Smorchkova1997,Andrearczyk2002}, i.e.: (a) significant low-field positive MR and (b) high-field negative MR. Qualitatively the same MR behavior was observed in both perpendicular and parallel fields \cite{Smorchkova1997,Andrearczyk2002}, which points to the importance of the giant spin splitting in scattering processes. In 3D DMS such positive MR behavior was explained \cite{Sawicki1986} by the destructive effect of giant exchange-induced spin splitting on the quantum correction to conductivity in the weakly localized regime. The observed increase of the resistivity with lowering temperature indicated the formation of magnetic polarons (MPs) \cite{Wolff1988,Dietl1982,Dietl1983}, ferromagnetic clouds of Mn ion spins that are polarized by a quasilocalized electron spin within its localization orbit. These MPs constitute centers of effective spin-disorder scattering that increases resistivity. Application of high magnetic field destroys MPs, and hence suppress scattering, which leads to the positive magnetoresistance. However, in modulation-doped QWs charged donors are removed from the conductive channel, thus the formation of MPs is less probable. Indeed in magneto-optical studies \cite{Smorchkova1998a,Smorchkova1998b,Kikkawa1998} no signatures of bound MP formation in DMS QWs have been found. It was debated that the existence of so-called free magnetic polarons --- spin-dressed carrier states --- may explain the experimental data \cite{Smorchkova1997,Smorchkova1998b}. Strong, high-field, negative MR in DMS QWs, observed close to the metal-insulator transition \cite{Jaroszynski2007}, was explained by the formation of ferromagnetic bubbles and corresponding electronic phase separation. At low temperatures, there is a competition between the insulating DMS phase characterized by antiferromagnetic interactions and the metallic phase which is governed by carrier-mediated ferromagnetic correlations \cite{Dietl2000}. Thus, ferromagnetic metallic bubbles are formed within a carrier-poor, magnetically disordered matrix. Such behavior resembles that of the colossal MR phenomenon observed in manganites \cite{Jin1994}. Low-field positive MR in DMS QWs is usually explained \cite{Smorchkova1996,Smorchkova1997,Smorchkova1998} by the quantum corrections to the conductivity originating from electron-electron interactions \cite{Lee1985,Altshuler1985}, modified by a giant spin splitting.

Magnetic impurities also affect transverse magnetotransport in DMS QWs, leading to the anomalous Hall effect (AHE) which occurs in all magnetic systems. Edwin Hall measured anomalously high "rotational coefficients" in nickel and cobalt \cite{Hall1881} almost a year after he discovered the Hall effect \cite{Hall1880}. Nowadays this phenomenon is known as AHE and its contribution to total Hall resistance is proportional to the magnetization of the studied material. Thus total Hall resistivity $\rho_{xy}$ is described by the empirical formula: $\rho_{xy}=R^{\text{H}}*B+R^{\text{AHE}}*M$, where $R^{\text{H}}$ is ordinary Hall coefficient related to carrier density and $R^{\text{AHE}}$ is anomalous Hall coefficient. However, the mechanism of AHE is rather complex and consists of three contributions: intrinsic, side jump, and skew-scattering \cite{Nagaosa2010}; the latter two mechanisms strongly depend on the strength of SO coupling. Since AHE occurs in magnetic materials, DMS-based QWs are a natural place for studying this effect in 2D. AHE contribution to the Hall effect was extracted in the case of paramagnetic $\text{Zn}_{1-x-y}\text{Cd}_{y}\text{Mn}_{x}\text{Se}$ QW in the Ref. \cite{Cumings2006}. A temperature-independent carrier density (and thus the ordinary Hall contribution) was calculated from the period of SdH oscillations, and magnetization was measured by magneto-luminescence. The temperature dependence of the extracted AHE contribution was found to follow the paramagnetic Brillouin-like magnetization of Mn ions (eq. (\ref{eqMn_mag})). It was found that the AHE coefficient $R^{\text{AHE}}\sim\rho_{xx}$, which was interpreted as the dominant skew-scattering mechanism despite the fact that the SO parameter in the ZnSe system is small \cite{Stern2006}.

Closely related to AHE is the spin Hall effect (SHE) \cite{Sinova2015}. While AHE is a phenomenon observed in systems with spin-imbalance, where spin-dependent scattering creates spatial charge separation, SHE occurs in systems with equally populated spin subbands, where spin-dependent scattering creates a spatial spin separation. Therefore, AHE and SHE are governed by the same mechanisms \cite{Sinova2015} and materials with a high SO constants are best suited for studying SHE. One of such systems is HgTe QW, in which a large value of the SO parameter was found \cite{Gui2004b}. In Ref. \cite{Bruene2010}, a high mobility HgTe/HgCdTe QW was used in a device geometry proposed in \cite{Hankiewicz2004}. In the applied experimental scheme the inverse SHE was used for the detection of a spin accumulation induced by SHE. HgTe QW was etched to form an H-shaped device, with two legs connected by a short channel. The electrical current applied to one of the legs generated a spin current through the connecting channel. The generated spin current, in turn, generated transverse voltage due to inverse SHE. This non-local voltage induced in the other leg of the device can dominate over local contributions if the separation between the legs is large enough. The device size was chosen to ensure operation in a quasiballistic regime. Gate voltage was used to control the strength of the Rashba SO coupling by variation of the electric field across the QW and the Fermi energy $E_F$ in the QW. Inverse SHE was found to be an order of magnitude higher in {\it p}-type regime with strong SO coupling than in weakly SO coupled {\it n}-type regime. Here it is worth noting that SHE was also detected using Kerr rotation measurements in 1.5~$\mu m$ thick epilayer of ZnSe \cite{Stern2006} despite the small value of the SO parameter in this material.

\subsubsection{Quantum Hall effect in 2D systems}

Integer QHE (IQHE) was observed by von Klitzing in 1980 \cite{Klitzing1980} as an appearance of plateaus in a Hall resistance $R_{xy}$ while varying the gate voltage in a Si-MOSFET. It was found that these plateaus in Hall resistance correspond to quantized values of $R_{xy}=\frac{h}{\nu e^2}$, where $\nu$ is an integer number, $h$ is the Planck constant, and $e$ is the electron charge. The precision of the QHE plateau quantization is 1 part per billion, which makes it a perfect resistance standard. Owing to the rapid progress in growing clean structures, soon after the discovery of IQHE the fractional QHE (FQHE) was discovered \cite{Tsui1982}. Each of the discoveries --- integer and fractional QHE --- was awarded the Nobel prize. QHEs were extensively studied mostly in GaAs/(Ga,Al)As QWs and recently in 2D materials, such as graphene.

QHE is observed in 2-dimensional systems, where carrier movement in $z$-direction (perpendicular to the 2D plane) is quantized. From quantum mechanics it is known that confinement of electron motion in a potential well leads to quantization of energy eigenvalues $E_i$ (called subbands) in the growth direction, while the motion in the 2D plane remains unaffected. Further, we will consider only the lowest subband ($i=0$) to be occupied. Two phenomena lay in the basis of IQHE, namely Landau quantization of energy levels in a strong magnetic fields, and the disorder-induced localization. Application of a magnetic field splits the 2DEG's energy levels into Landau levels (LLs) with energies: $E^N_c = \hbar\omega_c\left( N+\frac12\right)$. Here $\omega_c=eB/m^*$ is the cyclotron frequency, $\hbar$ --- reduced Planck constant, $m^*$ --- effective mass and $N$ is the LL index. Each LL is further split into two spin-resolved energy levels due to Zeeman effect. In wide-gap semiconductors the intrinsic Land{\'e} $g$-factor $g^*$ is relatively small (e.g., in ZnSe $g^*_{\text{ZnSe}}=+1.14$ \cite{Cavenett1981}; in CdTe $g^*_{\text{CdTe}}=-1.6$ \cite{Simmonds1982}) so the Zeeman splitting $E_z$ of LL is generally much smaller than the cyclotron gap. In a magnetic field $B$ 2DEG will split into the following spin-resolved LLs:
\begin{equation}
E_{N,\uparrow,\downarrow} = E^N_c + E_z = \hbar\omega_c\left(N+\frac12\right) \pm \frac12g^*\mu_BB,
\label{eqLL_spin}
\end{equation}
where $E^N_c$ is the cyclotron energy, $E_z$ is the Zeeman energy and $\pm\frac12$ stands for spin orientation $\uparrow$ and $\downarrow$.

In real samples, these discrete LLs are smeared into bands due to both thermal broadening and scattering. Thus, the resulting density of states in a high magnetic field is split into a sequence of broadened $\delta$-peaks at energy values corresponding to spin-split LLs. The states close to the center of the LL band are extended, while the states in the tails of LLs are localized and do not contribute to transport. Only the extended states can carry current at zero temperature. When these extended states overlap the Hall resistance $R_{xy}$ is not quantized and the longitudinal resistance $R_{xx}$ exhibits SdH oscillations due to the oscillating density of states. With increasing of a magnetic field this overlap diminishes. If $E_F$ lays between LL bands, $R_{xy}$ exhibits plateaus at the quantized values $\frac{h}{\nu e^2}$, where $\nu$ (called filling factor) corresponds to number of LL bands below $E_F$, and $R_{xx}=0$. Whenever $E_F$ passes the center of a LL band, $R_{xy}$ makes a transition between plateaus and $R_{xx}\neq0$ during this transition.

A widely used description of electrical transport in the QHE regime is the edge channel picture \cite{Halperin1982} and the Landauer-B{\"u}ttiker formalism \cite{Buettiker1988}. Within this picture, 1D edge channels are formed at the intersection of $E_F$ with LLs bending at the physical edge of the sample. This can be intuitively understood by treating the exterior of the sample as even higher energy barriers than those which form the QW. As a result, energy levels in the exterior of the sample lie much higher than those inside QW and LLs inside the QW have to smoothly level out to their outside values. The number of edge channels is equal to the number of occupied bulk LLs, and the conductance of each edge channel is $G = \frac{e^2}{h}$, independent of its length. They are also chiral, which means that their direction is determined by the direction of the applied magnetic field.

QHE in semiconductor heterostructures is usually observed at helium temperatures and high magnetic fields when the thermal broadening of LLS is lower than the cyclotron energy. Hence, lower $m^*$ (and consequently the higher $E^N_c$) reduce requirements for the temperature and magnetic field. This is important for the metrology applications since it makes QHE achievable at lower magnetic fields and higher temperatures. QHE at room temperature and 30-45~T magnetic field was observed in graphene, because of the huge LL splitting for massless Dirac fermions \cite{Novoselov2007}. The Dirac semimetal phase can be also achieved in chalcogenide HgTe/HgCdTe QWs for a particular critical QW thickness $d_c$ \cite{Buettner2011}. QHE states with large energy gaps were indeed observed in QWs with thicknesses close to $d_c$ \cite{Kozlov2014,Khouri2016}. It was argued that further optimization can be obtained with the use of a strained HgTe QW, which would result in a better QHE performance. Thus, achieving precisely quantized plateaus would be possible at magnetic fields accessible with permanent magnets \cite{Yahniuk2019}.

In the DMS-based QW the LL energies are modified because of the presence of the Mn-subsystem. The spin-splitting term in eq. (\ref{eqLL_spin}) in this case contains an additional exchange term $E_{exch}$, which is proportional to the magnetization of the Mn-subsystem (eq. (\ref{eqMn_mag})):
\begin{multline}
E_{N,\uparrow,\downarrow} = E^N_c + E_z + E_{exch} = \\ = \hbar\omega_c\left(N+\frac12\right) \pm \frac12\left(g^*\mu_BB + x_{\text{eff}}E_{sd}S\mathfrak{B}_s\left(B,T\right)\right),
\label{eqLL_exch}
\end{multline}
where $E_{sd}$ stands for s-d exchange energy \cite{Furdyna1988} and the Brillouin function $\mathfrak{B}_s\left(B,T\right)$ has the same argument as in eq.(\ref{eqMn_mag}). Introducing an effective $g$-factor $g_{\text{eff}}$:
\begin{equation}
g_{\text{eff}} = g^* + \frac{x_{\text{eff}}E_{sd}S\mathfrak{B}_s\left(B,T\right)}{\mu_BB},
\label{eqgeff}
\end{equation}
allows to rewrite eq. (\ref{eqLL_exch}) in the same form as eq. (\ref{eqLL_spin}). 

\begin{figure}[t]
\centering\includegraphics[width=0.75\columnwidth]{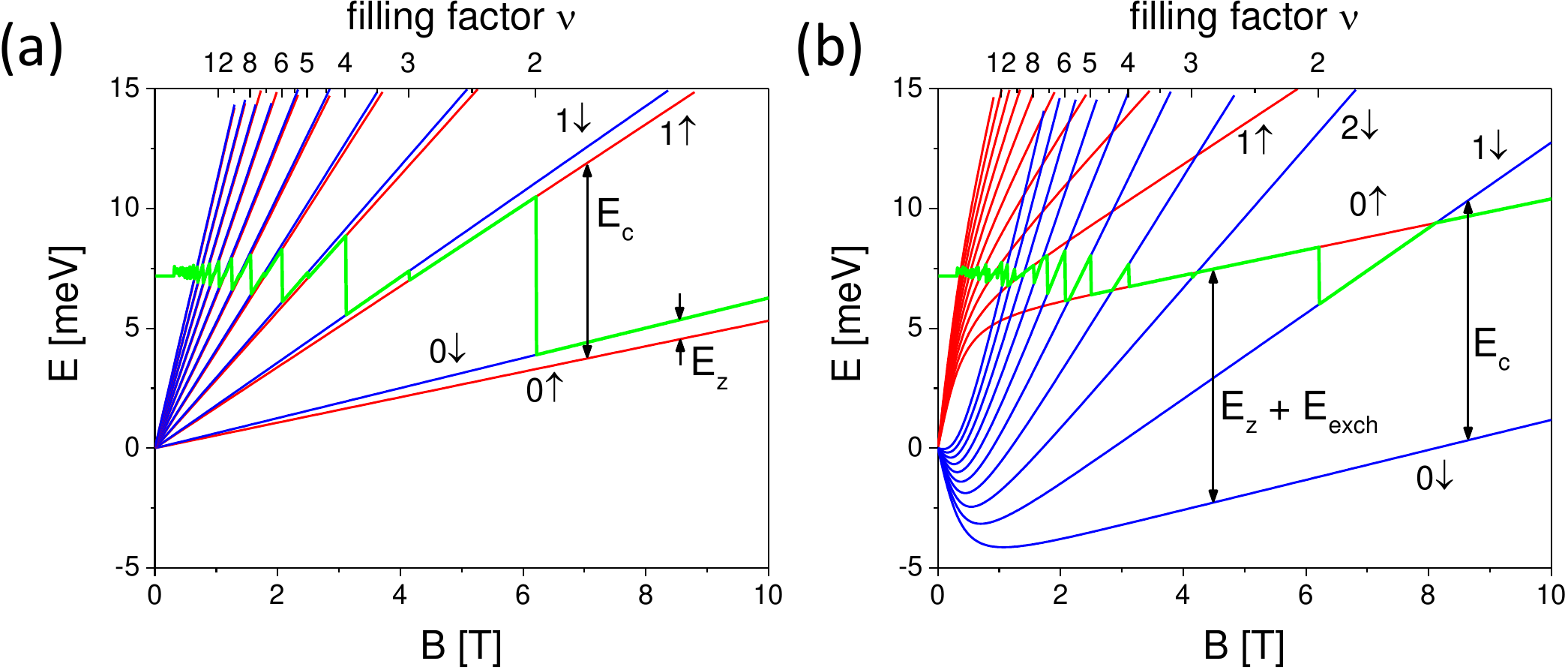}
\vspace{0in}
\caption{Calculated LL fan diagram (blue and red lines denote different spin orientations) and chemical potential (Fermi energy, green line) for (a) QW without paramagnetic impurities and (b) QW with paramagnetic impurities.Calculations were performed for carrier density $3\cdot10^{11}~cm^{-2}$; LLs were calculated according eq. (\ref{eqLL_exch}) using $E_{sd}=220~meV$ for (Cd,Mn)Te \cite{Furdyna1988} and $x_{\text{eff}}=1.85\%$. Here, for the simplicity, we neglected corrections which arise from electron-electron interactions \cite{Kunc2010}.}
\label{fLLfan}
\end{figure}

The exchange term $E_{exch}$ in DMS QWs leads to a strong modification of the LL chart, see Fig.\ref{fLLfan}. First of all, it can be seen that for $x_{\text{eff}}>1\%$ the spin-splitting energy $E_z+E_{exch}$ is comparable to the cyclotron energy $\hbar\omega_c$. In conventional, non-magnetic QWs such condition is usually achieved by tilting 2DEG by large angles in high magnetic fields \cite{Eisenstein1990,DePoortere2000}, since $E_z$ depends on the total $B$ and $E^N_c$ depends only on the perpendicular component $B_\perp$ of $B$. In Mn-doped chalcogenide QWs it can be achieved solely by controlling Mn doping \cite{Smorchkova1997,Wojtowicz2000,Jaroszynski2002,Teran2002} or by gate voltage in asymmetrically doped QWs \cite{Kazakov2016,Kazakov2017}. In both cases, $B$ is perpendicular to the 2DEG plane, which allows to avoid using very high magnetic fields, that are usually applied to keep $B_{\perp}$ large enough to maintain QHE quantization. Also, it should be noted that LL spectrum can be further distorted in CdTe QWs by the correction to the spin gap between fully occupied LLs \cite{Kunc2010}.

\subsubsection{Modification of Shubnikov-de Haas oscillations}

The first studies of quantum transport in chalcogenide DMS 2DGEs were actually performed not on QW structures but on the inversion layers created on the surface of (Hg,Mn)Te bulk crystals \cite{Grabecki1984a} or naturally occurred in the grain boundaries of (Hg,Cd,Mn)Te \cite{Grabecki1984,Grabecki1987,Grabecki1993,Grabecki1995}. Ref. \cite{Grabecki1984a} reported probably the first evidence of the modification of SdH oscillations in a 2DEG due to the presence of magnetic impurities, i.e., strong temperature dependence of the positions SdH oscillations. Also, a QHE in DMS was observed in grain boundaries in bulk (Hg,Cd,Mn)Te crystals \cite{Grabecki1987,Grabecki1993}. Well developed QHE plateus were reported in a single isolated grain boundary. Analysis of activation energies, which corresponded to different filling factors, also confirmed the modification of LLs in the presence of magnetic impurities.

As mentioned above, SdH oscillations in MR have been already observed in the first obtained chalcogenide QWs. Soon after that, with increasing quality of the 2D systems grown, QHE with well developed plateaus in $R_{xy}$ was observed in both nonmagnetic and magnetic QWs \cite{Smorchkova1997,Karczewski1998,Jaroszynski2000}. Already the first measurements revealed a modification of the QHE states sequence in magnetic QWs compared to non-magnetic ones \cite{Smorchkova1997}. Only even number plateaus ($\nu$ = 2, 4, 6), with the only exception of $\nu = 1$, were observed up to 10~T in nonmagnetic $\text{Zn}_{0.8}\text{Cd}_{0.2}\text{Se}/\text{ZnSe}$ QW, indicating that spin splitting is rather small in the studied system. On the other hand, both even and odd filling factors were clearly observed in DMS-based QWs starting from 2~T. This is a clear manifestation of spin splitting enhancement by the s-d exchange term ($E_{exch}$ in eq. (\ref{eqLL_exch})). Moreover, other anomalies in the SdH pattern were observed in DMS-based QWs, such as additional resistance peaks and deviations from the periodicity of SdH versus 1/B \cite{Jaroszynski2000,Knobel2002}. The anomalies observed in magnetotransport data agreed well with magnetization measurements, i.e., with the de Haas-van Alphen oscillation pattern obtained using cantilever magnetometry \cite{Harris2001}. The above mentioned anomalies in SdH oscillations were thoroughly studied in later works \cite{Teran2002,Jaroszynski2002} which have shown that they also stem from giant spin splitting, which modifies the energy spectrum of LLs (see Fig.\ref{fLLfan}).

\begin{figure}[t]
\centering\includegraphics[width=0.75\columnwidth]{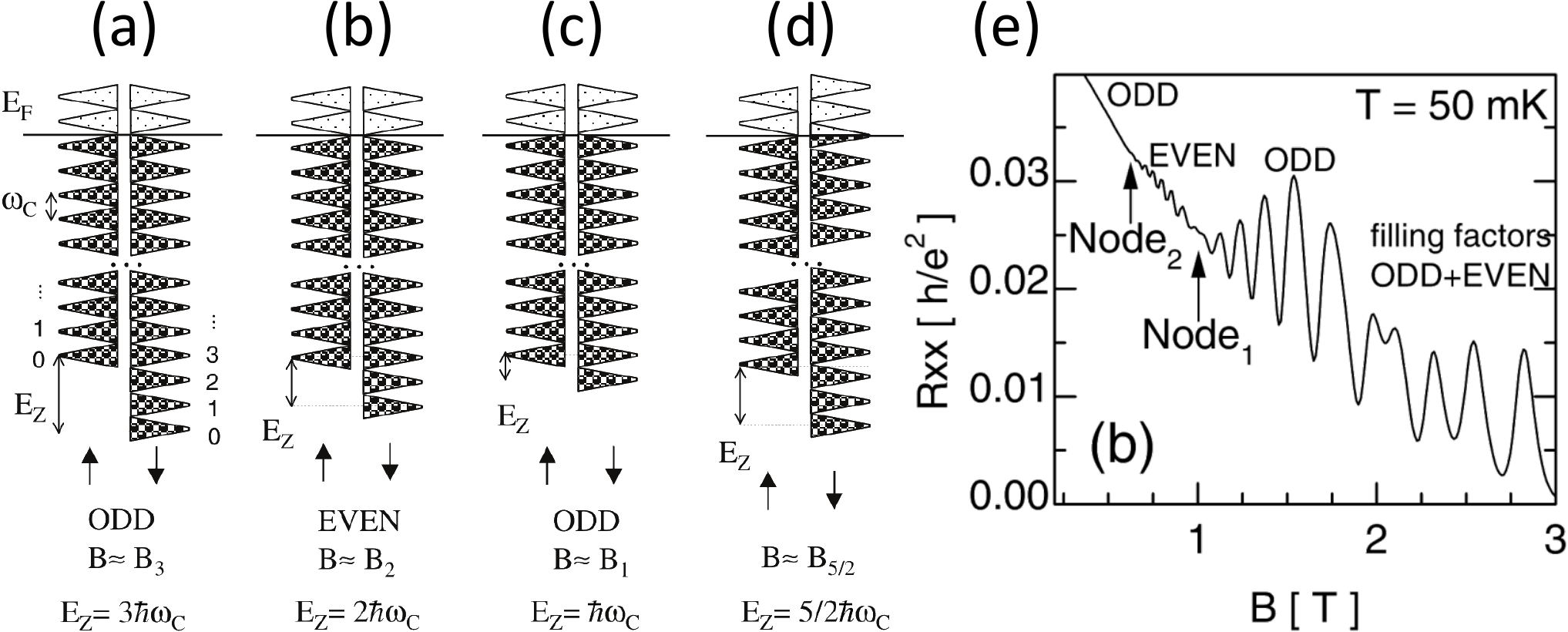}
\vspace{0in}
\caption{(a-c) Possible arrangements of spin-up and spin-down LLs when Zeeman energy is a multiple of the cyclotron energy. In these cases, the Fermi energy $E_F$ lies in the gap of the density of states which corresponds to a minimum in $R_{xx}$. (d) Arrangements of spin-up and spin-down LLs for a magnetic field corresponding to the case when the Zeeman energy is half-integer of the cyclotron energy (e.g. $B_{5/2}$). In this case, there is no gap in the density of states for any filling factor and a node in the beating pattern of $R_{xx}$ is observed. (e) Beating pattern in SdH oscillations in the low magnetic field region. Adapted with permission from Ref. \cite{Teran2002}.}
\label{fbeating}
\end{figure}

At low Mn concentrations, when $E_{exch}$ (eq. (\ref{eqLL_exch})) is rather small and the carrier mobility is relatively high, SdH oscillations are well resolved already at low magnetic fields ($B\leq2~T$). A beating pattern in SdH oscillations is present in such samples \cite{Teran2002,Gui2004a,Kunc2015}. At low magnetic fields the spin-splitting energy is larger than the cyclotron energy and the actual arrangement of LLs depends on the applied magnetic field (see Fig.\ref{fbeating}a-d). In the case when the spin-splitting energy $E_z+E_{exch}$ is an integer multiplier of the cyclotron energy $\hbar\omega_c$, LLs from both spin subbands coincide (Fig.\ref{fbeating}a-c) and a maximum in the SdH amplitude is observed. When $E_z+E_{exch}$ is a half-integer multiplier of $\hbar\omega_c$, the centers of LLs from one spin subband are aligned between LLs from the opposite spin subband (Fig.\ref{fbeating}d). Therefore, there are no oscillations in the density of states and a node in the beating pattern is observed. Usually, such beating is visible at low fields $B<1-2~T$, when QHE plateaus are not well resolved. Furthermore, dips in SdH oscillations between nodes correspond to an alternating sequence of even and odd filling factors, which depends on the difference between the number of filled LLs for opposite spins, see Fig.\ref{fbeating}a-c. Due to the fact that at such low fields and elevated temperatures the Brillouin function contained in the exchange term (eq.\ref{eqLL_exch}) is not yet saturated, $T_{AF}$ can be obtained from the temperature dependence of the node positions. Analyses of SdH beating patterns provide slightly different values for $T_{AF}$: 180~mK \cite{Teran2002} and 40~mK \cite{Kunc2015} in (Cd,Mn)Te QWs with $x_{\text{Mn}}=0.3\%$, and 2.6$\pm$0.5~K \cite{Gui2004a} in a HgMnTe QW with $x_{\text{Mn}}=2\%$. 

Similar behavior, namely a low-field SdH beating pattern, can also arise in the systems where spin splitting is induced solely by SO coupling \cite{Das1989}. In a QW structure SO effects are mainly governed by structure asymmetry, e.g., produced by gating. In the case of the (Hg,Mn)Te QW, SdH beating was used to separate contributions to the spin splitting arising from the gate-dependent Rashba effect and temperature-dependent spin splitting caused by exchange interactions \cite{Gui2004a}.

In Ref. \cite{Kunc2015} the exchange term from eq. (\ref{eqLL_exch}) was incorporated into the Lifshitz-Kosevich formalism in a mean-field approach in order to describe quantitatively the low-field SdH oscillations. Beside the s-d exchange and Zeeman contributions to the spin-splitting energy, additional terms were considered, such as the contribution of antiferromagnetic interactions within pair clusters of Mn atoms \cite{Aggarwal1985} and contribution of electron-electron interactions \cite{Kunc2010}. Considering these terms allowed a better fit of the detailed temperature and field dependencies of node positions as well as determination of cluster formation probability, which was found to be 20~\% (for $x_{Mn}\sim0.3$~\%). The obtained temperature and field dependencies of LL broadening in (Cd,Mn)Te QWs suggested that there is an additional source of LL broadening, caused by inhomogeneity of Mn distribution on the local scale. However, comparison of SdH oscillations in CdTe and (Cd,Mn)Te QWs gives approximately the same values for the total LL broadening in the two systems, thus LL broadening originating from other sources should be smaller in (Cd,Mn)Te QW than in CdTe QW. This can be explained by the suppression of spin-flip scattering between LLs in (Cd,Mn)Te. Indeed, in CdTe QW LLs with opposite spins and the {\it same} $N$ are close to each other. On the other hand, in (Cd,Mn)Te the giant spin splitting brings into coincidence LLs with {\it different} indexes $N$ and spin orientations. Thus, spin-flip scattering would occur between LLs with different orbital quantum numbers, which may affect the scattering rate between them. Altogether, the proposed modifications allowed achieving a quite good fit to the exchange-induced SdH beating pattern at low $B$.

\subsubsection{Quantum Hall ferromagnetic transition}

At high magnetic fields, as shown in Fig.\ref{fLLfan}b, LLs with different spin orientations cross each other. Thus, it is possible to have 2DEG being partially or fully spin-polarized. Transitions between QHE states with different spin polarization but the same $\nu$ are called quantum Hall ferromagnetic (QHF) transitions \cite{Girvin2000,DePoortere2000,Jaroszynski2002}. Such transitions were observed both in the integer \cite{DePoortere2000,Jaroszynski2002} and in the fractional \cite{Eisenstein1990,Betthausen2014} QHE regime. The QHF transition is accompanied by an additional peak in the longitudinal resistance $R_{xx}$. The properties of QHF transitions in (Cd,Mn)Te QWs were extensively studied in the 2000's \cite{Jaroszynski2002,Teran2010}, for a popular description see Ref. \cite{Girvin2000} and for a short review see Ref. \cite{Jaroszynski2010}.

When two LLs are brought into coincidence at a sufficiently low temperature they split into one fully occupied and one empty LL, provided that the energy gain is larger than the LL band width. This happens below a certain critical temperature $T_c$, since LL widths decrease with temperature. Below $T_c$ the Fermi energy lies in the energy gap between two LLs and at the lowest temperature, when the thermal broadening of LLs is lower than the energy gap, the QHE state with $R_{xx}=0$ and $R_{xy}=\frac{1}{n}\frac{h}{e^2}$ recovers. In magnetic fields on either side of the crossing (in Fig.\ref{fLLfan}b, e.g., the $\ket{0\uparrow}$ state crosses the $\ket{1\downarrow}$ state at $B^*=8~T$) 2DEG has different spin polarizations of the top filled LL.

\begin{figure}[t]
\centering\includegraphics[width=0.75\columnwidth]{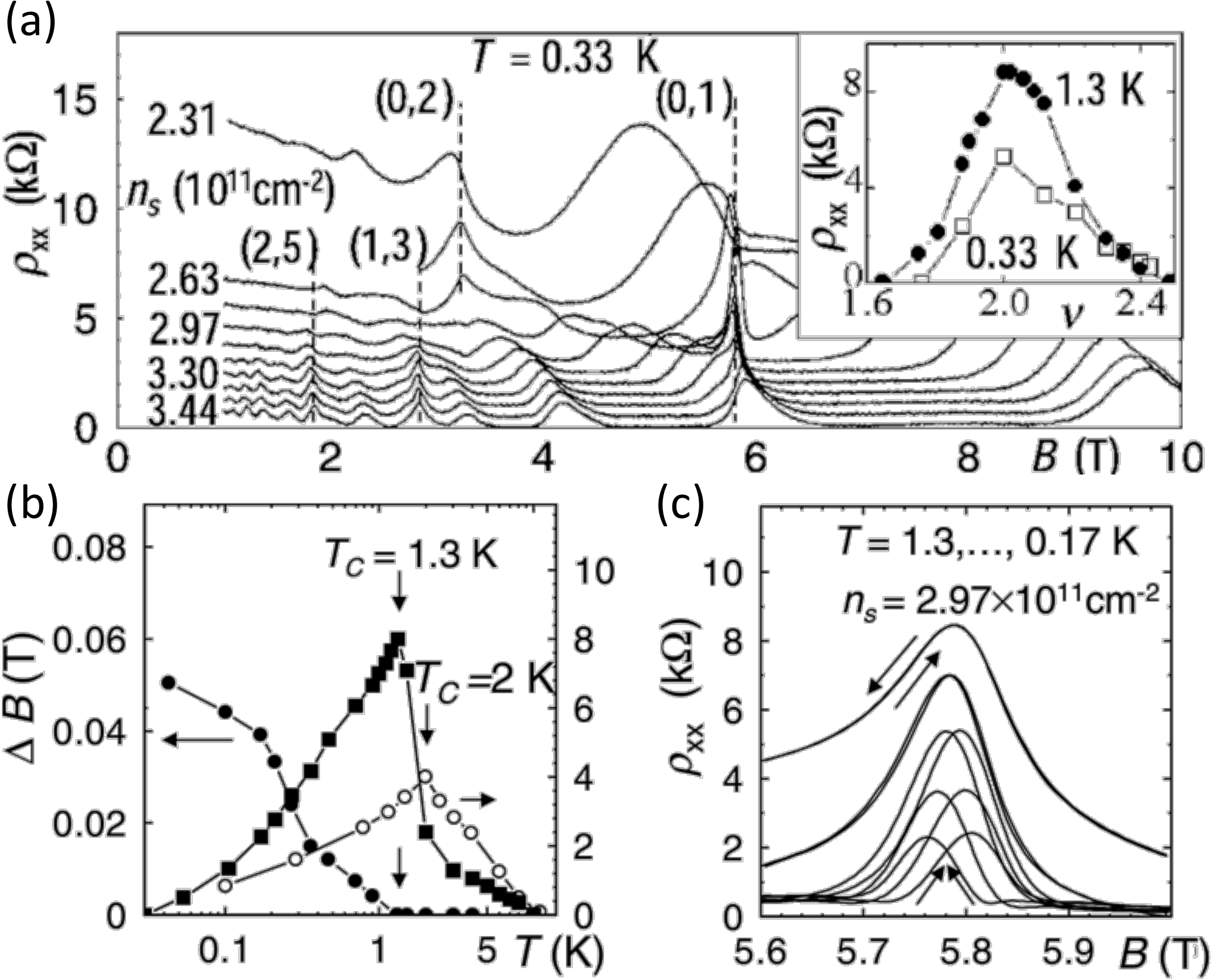}
\vspace{0in}
\caption{(a) Longitudinal resistance $\rho_{xx}$ measured at 0.33~K for carrier density in the range of $2.31-3.44~10^{11} cm^{-2}$. Traces are shifted vertically for clarity. Dashed lines mark additional resistance spikes associated with LL crossings. Numbers in brackets indicate indexes of crossing LLs. (b) Temperature dependence of the spike height for two different carrier densities ($n=2.97$ (squares) and $2.54~10^{11}~cm^{-2}$ (open circles)). Full circles show the temperature variation of the difference between peak positions $\Delta B$ obtained for opposite field sweep directions (presented in (c) for different temperatures). Adapted with permission from Ref. \cite{Jaroszynski2002}.}
\label{fqhf}
\end{figure}

The behavior of the peak in $R_{xx}$ which accompanies the QHF transition at finite non-zero temperature was studied in details in Ref. \cite{Jaroszynski2002}. The following features of the resistance peak were reported:
\begin{enumerate} 
\item Dependence of the peak amplitude on the filling factor $\nu$ with a maximum at $\nu=2\pm0.02$ (see Fig.\ref{fqhf}a);
\item Hysteretic behavior of the peak position, which manifests itself as a change of the crossing field $B^*$ under an upward and downward magnetic field sweep (see Fig.\ref{fqhf}c);
\item Strong temperature dependence: the peak amplitude increases with the temperature reaching its maximum value at $T_c$ and then decreases with a further increase of temperature, while the hysteresis in the peak position gradually decreases with temperature and disappears at $T_c$ (see Fig.\ref{fqhf}b).
\end{enumerate}

While Ref. \cite{Jaroszynski2002} points out that one-electron anti-crossing effects driven by SO interactions are of minor importance, in Ref. \cite{Kazakov2016} the temperature dependence of the peak maximum was interpreted as SO-induced anti-crossing of LLs. Numerical calculations of the SO-induced energy gap between $\ket{1\downarrow}$ and $\ket{0\uparrow}$ LLs yielded value in quantitative agreement with experimental data. A more thorough experimental study of the anti-crossing gaps is presented in Ref. \cite{Teran2010}, where energy gaps of QHF transitions for different filling factors were reported. However, to achieve LL crossings at different filling factors the sample had to be tilted. It was found that the energy gap depends on the filling factor --- e.g., an anti-crossing energy gap $\Delta_S$ was closed for filling factors $\nu=5$ and $7/2$ and opened for $\nu=3, 5/2$ and $3/2$. These results were explained assuming that $\Delta_S$ is modified by disorder mediated electron-electron interactions. The effect of disorder was also noticed in Ref. \cite{Kazakov2017} for the filling factor $\nu=2$. It is worth to note that a peak in $R_{xx}$, which would correspond to a crossing between states $\ket{0\downarrow}$ and $\ket{0\uparrow}$, was not observed in transport measurements. A possible reason for that is the high value of the energy gap for this anti-crossing --- approximately 10~K, according to the calculations of SO-induced gap \cite{Kazakov2016}.

QHF transitions have also consequences for the configuration of edge channels in the vicinity of the sample edge. The bulk filling factor in the QHE regime reduces in a step-like manner when approaching the sample edge \cite{Chklovskii1992}. In a DMS-based QW, during the QHF transition, the arrangement of LLs changes with the magnetic field (Fig.\ref{fLLfan}b). A corresponding change of the edge channel arrangement is hence expected. A theoretical study \cite{Rijkels1994} has shown that edge channel crossings are soliton-like, one-dimensional, Bloch magnetic domain walls, which can be induced on-demand under certain conditions. Recently, such phenomena were studied experimentally through observing magnetoconductance resonances \cite{Bobko2019}. Experimental data, interpreted in terms of crossings of edge channels originating from different LLs, agreed well with the theoretical picture, though not completely.

\subsubsection{Fractional quantum Hall effect in DMS QWs}

As it was said above, if the samples are higher quality, i.e. higher mobility, plateaus with fractional filling factor $\nu$ emerge between plateaus in $R_{xy}$ with an integer $\nu$. In the experiment \cite{Tsui1982} on a GaAs/AlGaAs QW a QHE plateau with $\nu=\frac13$ was observed, which was quite unexpected since there is no energy gap below $\nu=1$ in IQHE picture. Explanation of this effect required accounting electron-electron interactions, which were excluded before. A widely accepted theory, which describes FQHE, is a composite fermion (CF) picture \cite{Jain1989,Jain2015}. Within the CF description, each electron attaches an even number ($2p$) of quantized vortices, and electronic states are transformed into composite fermion states. CFs are weakly interacting quasi-particles which are moving in reduced magnetic field: $B^*=B-2pn\phi_0$, where $n$ is carrier density, and $\phi_0=\frac{h}{e}$ is flux quantum. CFs are forming their own LLs in the effective $B^*$, which are called $\Lambda$ levels or $\Lambda$Ls. Filling factor for CFs $\nu^*=\frac{n\phi_0}{|B^*|}$ is related to electron filling factor: $\nu = \frac{\nu^*}{2p\nu^*\pm1}$, where minus sign corresponds to a case when $B^*$ is opposite to $B$.

\begin{figure}[t]
\centering\includegraphics[width=0.75\columnwidth]{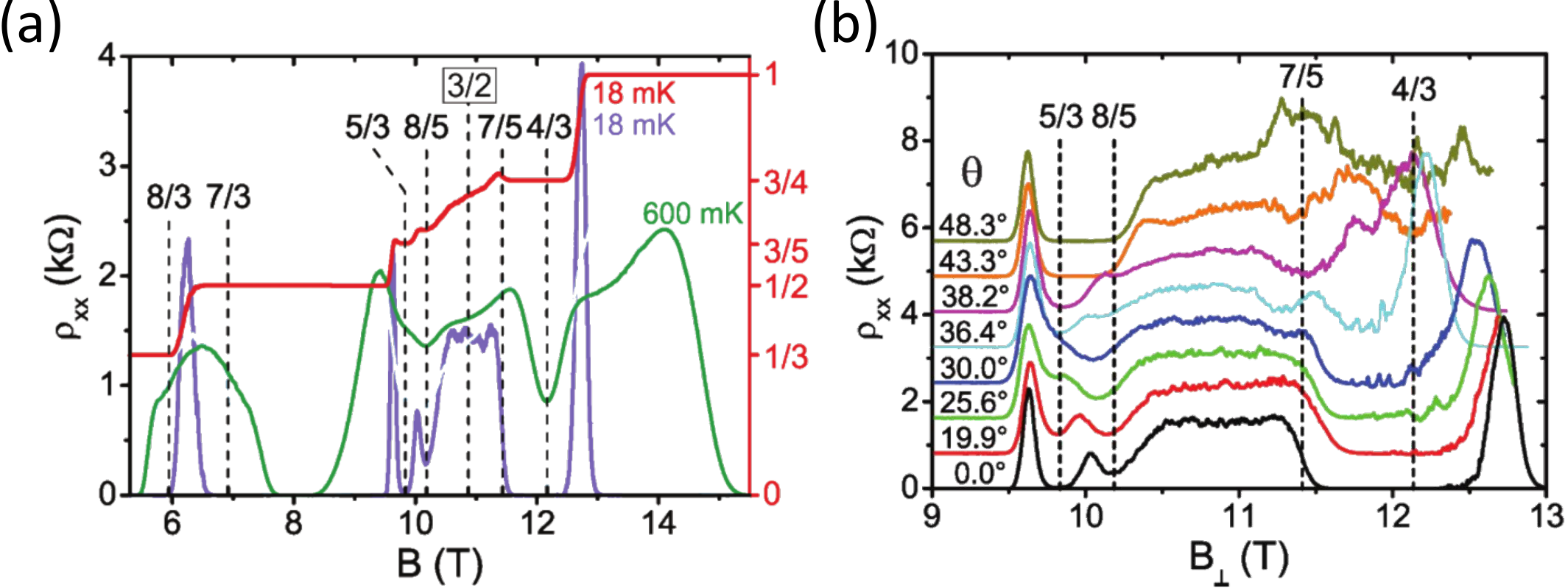}
\vspace{0in}
\caption{(a) Longitudinal and Hall resistances in a high mobility (Cd,Mn)Te QW in perpendicular magnetic field. Fractional filling factors are indicated by dashed lines. (b) Angular evolution of FQHE states in the vicinity of $\nu=3/2$. Remarkably, $5/3$, $4/3$, and $8/5$ FQHE states first disappear at intermediate tilt angles and then reappear at larger angles. In contrast, the $\rho_{xx}$ minimum corresponding to the $7/5$ state can be only observed in the intermediate range at tilt angles around $38^{\circ}$. Adapted with permission from Ref. \cite{Betthausen2014}.}
\label{ffqhe}
\end{figure}

Recent progress in the growth of CdTe/CdMgTe heterostructures made it possible to observe FQHE not only in non-magnetic CdTe QWs \cite{Piot2010} but also in DMS (Cd,Mn)Te QWs \cite{Betthausen2014} (see Fig.\ref{ffqhe}a). Fractional filling factors were observed in CdTe-based 2DEG samples with slightly lower mobilities than those needed in GaAs QWs to observe FQHE. The quantum scattering time $\tau^q$, which is considered a predictor for the strength of FQHE states \cite{DasSarma2014}, turns out to be higher in both non-magnetic CdTe and magnetic (Cd,Mn)Te QWs ($\tau^q_{\text{CdTe}}\approx3\pm0.3~ps$ \cite{Piot2010,Betthausen2014}) than in GaAs QWs of same mobility. Another important difference between FQHE states in GaAs and CdTe QWs is the different Zeeman energy scale. Due to the fact that the $g$-factor for CFs, $g^{CF}$, in CdTe is $\sim-1.99$ \cite{Betthausen2014}, which is larger than in GaAs ($g_{\text{GaAs}}^{CF}=-0.61$ \cite{Du1995}), FQHE states were found to be completely spin polarized \cite{Piot2010}. The incorporation of paramagnetic Mn ions in (Cd,Mn)Te QWs leads to crossing of $\Lambda$Ls, which can be described by equation (\ref{eqLL_exch}), modified for the CF case:
\begin{equation}
E_{N,\uparrow,\downarrow} = \hbar\omega_c^{CF}\left(N+\frac12\right) \pm \frac12\left(g^{CF}\mu_BB+ x_{\text{eff}}E_{sd}S\mathfrak{B}_s\left(B,T\right)\right).
\label{LL_CF}
\end{equation}
Here almost all notations are the same as in equation (\ref{eqLL_exch}), except $g^*$ for electrons is replaced by the $g^{CF}$ for CFs, N being now the index of CF $\Lambda$L, and $\omega_c^{CF}\sim\frac{e^2}{\epsilon l_B}$ the CF cyclotron energy with $\epsilon$ --- the dielectric constant and $l_B=\sqrt{\hbar c/eB}$ --- the magnetic length. Addition of the exchange term for CFs made it possible to reproduce the observed angular dependence of the energy gap for $\nu=\frac53$, i.e., closing and re-opening of the gap (see Fig.\ref{ffqhe}b), which corresponds to a change of spin polarization.

An important result of the observation of FQHE in DMS QWs is that incorporation of a relatively large number of magnetic impurities ($x_{\text{eff}}=0.3\%$) does not inhibit the formation of FQHE states. However, only fractional states with relatively large energy gaps ($\nu=\frac23, \frac43, \frac53$) were observed in experiments \cite{Piot2010,Betthausen2014}. (Cd,Mn)Te QWs with even higher quality are needed to study how modification of the CF spin splitting energy affects the versatile FQHE phenomena \cite{Jain2015}.

\subsubsection{Magnetotransport in wide HgTe QWs}

In a wide HgTe (013) QW with $d=20.5$~nm a non-linear Hall resistance was observed at low magnetic fields \cite{Kvon2008}. It was interpreted as a coexistence of two types of carriers --- electrons and holes. It was argued that in such a wide QW the conduction and valence bands are formed by the H1 and H2 bands, respectively. The latter has a local maximum away from the center of the Brillouin zone and there is an overlap with the conduction band, which minimum is at $k=0$. The overlap is of the order of 2-5~meV and depends on a QW thickness. When $E_F$ coincides with this overlap mixed-type conductivity is observed. Thus, a wide HgTe QW constitutes a 2D electron-hole system. Calculations have shown that the strain (caused by the lattice mismatch between HgTe and CdTe) plays a crucial role in the formation of the overlap between conduction and valence bands \cite{Kvon2011}. Extensive transport measurements \cite{Minkov2013} later confirmed this picture qualitatively, however, quantitatively they gave different results for electron and hole effective masses as well as the position of the local maximum for the H2 band in {\it k}-space.

QHE was measured in the region of mixed-type conductivity \cite{Gusev2010}. It was found that though there was no plateau in $R_{xy}$ at high fields, there was a plateau for $\nu=0$ in $\sigma_{xy}$, accompanied with non-zero $\sigma_{xx}$. However, this state did not exhibit the usual activation behavior with $\rho(T)\sim e^{\Delta/2kT}$. Moreover, the longitudinal MR was strong, while the Hall resistance was 0. It was claimed, that the origin of this QHE state is the formation of electron-hole "snake states" at the boundary between electron and hole puddles, caused by smooth fluctuations of the local filling factor. These "snake states" are similar to those which arise in the systems with inhomogeneous magnetic fields \cite{Nogaret2010}.

\subsubsection{Magnetotransport in IV-VI QWs}

In the past, there were several studies of magnetotransport in PbTe/PbEuTe QWs. In Ref. \cite{Peres2014} SO coupling was studied by means of the weak antilocalization effect. Weak antilocalization is a result of quantum interference, which manifests itself as a positive correction to the low-field conductivity \cite{Hikami1980,Bergmann1984,Lee1985}. Values of the SO gap $\Delta_{SO}$ were found to be in the range from $0.17$ to $0.60$~meV, depending on QW thickness. These values are comparable to those found in III-V semiconductor QWs. The Rashba term was dominant in the SO splitting.

QHE was also observed in IV-VI QWs. The main problem in the studies of QHE in these QWs is the significant parallel conductance \cite{Kolwas2013} resulting from growth on $\text{BaF}_2$ substrates, which is responsible  for non-local signals and prevents observation of precise QHE quantization. In early studies \cite{Olver1994} neither quantized plateaus in $R_{xy}$, nor $R_{xx}=0$ were observed in the field range up to 10~T. The problem of parallel conductance can be solved in structures with higher europium content in the barriers. QHE quantization was achieved in later experiments \cite{Chitta2005} performed on such structures, though at much higher magnetic fields ($B>10~T$). In the cited study an unusual sequence of QHE states was observed ($\nu=15, 11, 10, 6, 5, 4$), which was explained by assuming that there are three occupied electrical subbands in the QW, two originating from the longitudinal valley and one from the oblique valleys.

\subsection{DMS QW in inhomogeneous magnetic fields}

Now we turn to the behavior of 2DEG in spatially non-uniform magnetic fields. The dynamics of two-dimensional electrons in microscopically inhomogeneous magnetic fields exhibits some peculiarities and differs from electron motion in a spatially uniform magnetic field \cite{Nogaret2010}. There are several ways how an inhomogeneous magnetic field can be created. First of all, a microstructure of magnetic elements can create an inhomogeneous stray field in the plane of electron motion. Another way is to place a superconductor on top of the QW. Superconducting elements made of type-II superconductors are penetrated by the magnetic fields in the form of flux tubes if the applied magnetic field is higher than the first critical field $H_{c1}$. These tubes modulate the magnetic field in the QW plane in which electrons are moving \cite{Bending1990,Geim1992}. Transport properties of such hybrid structures with periodic inhomogeneities \cite{Gerhardts1989,Ye1995} were extensively studied for non-magnetic 2DEGs in the 90s. Weak 1D electrostatic periodic modulation of the QW potential has led to the discovery of a new type of MR oscillations, called {\it commensurability oscillations} \cite{Gerhardts1989}. The influence of 1D periodic modulation on 2D electron motion was treated with perturbation theory. Within this approach \cite{Peeters1993}, LLs broaden into bands whose widths depend on the applied uniform magnetic field. Due to the finite bandwidth a non-zero contribution to transverse conductivity appears which leads to an increase of longitudinal resistance. Every time the Landau bands at the Fermi energy becomes flat, the transverse conductivity vanishes and $R_{xx}$ exhibits minimum. In the case of DMS QWs, magnetic field inhomogeneity can lead to new topological phenomena \cite{Bruno2004,Berciu2005} and to new applications in spintronics \cite{Betthausen2012a}.

Experimental studies of hybrid structures involving DMS QWs are rather scarce and limited mainly to optical studies \cite{Crowell1997,Kossut2001,Kudelski2001,Cywinski2002}. It was shown in Ref. \cite{Crowell1997} that photoluminescence spectra changes after deposition of an iron film on the cap layer of the $\text{Zn}(\text{Mn})\text{Se}/\text{Zn}_{0.80}\text{Cd}_{0.20}\text{Se}$ QW. However, it was claimed that these changes are caused by the strain induced by the iron layer rather than by stray fields. The absence of the observable effects due to stray fields was explained by the difference between the size of the laser spot in the experiment ($\sim10~\mu m$) and the size of the edge region of the deposited magnetic layer in which the magnetic field drops off ($\sim100~nm$). Results of local photoluminescence measurements \cite{Kudelski2001,Cywinski2002} were interpreted in terms of excitons trapped locally in the potential minimum produced by the fringe magnetic field of the micrometer size iron island.

\begin{figure}[t]
\centering\includegraphics[width=0.75\columnwidth]{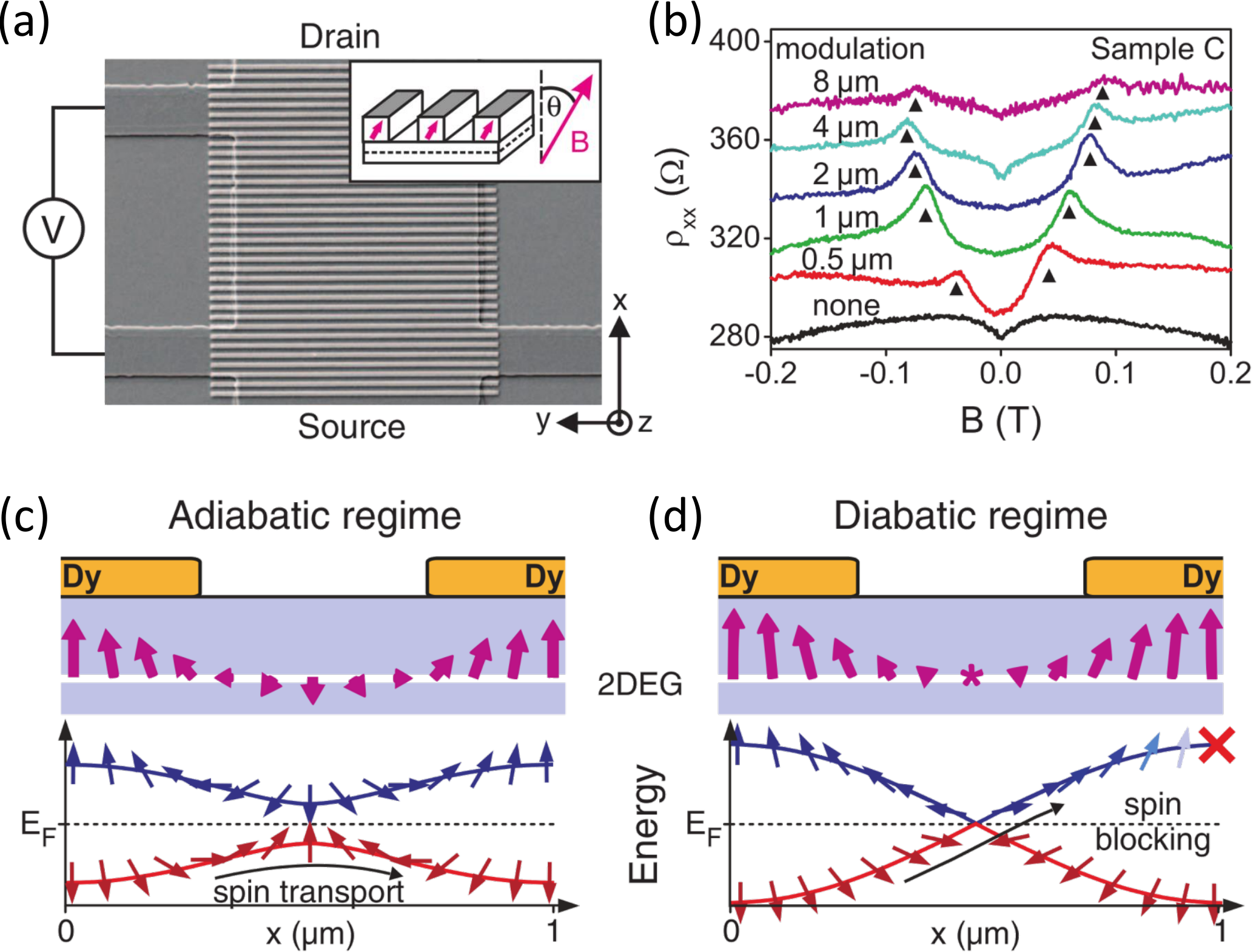}
\vspace{0in}
\caption{(a) Sample fabricated in the form of a hall-bar with Dy stripes grating on its surface. (b) MR in studied devices exhibits pronounced maximum at a particular field for different spacing between stripes. (c) Electron spin rotation in the adiabatic regime; the spin direction follows that of the slowly varying total magnetic field, which is a superposition of the externally applied field and stray fields from magnetized Dy stripes. Since the external field does not compensate completely the stray fields anywhere on the path of an electron, the adiabatic theorem applies and the electron remains in the ground state during the motion (“on” state). (d) Electron spin dynamics in the diabatic regime. In the middle between Dy stripes the external field completely compensates the stray fields, thus creating a discontinuity in the direction of the total field. Electrons moving from the source to drain cannot adjust to such a sudden change of its spin direction in order to remain at the lowest energy level, and this leads to backscattering and increase of the device resistance (“off” state). Adapted with permission from Ref. \cite{Betthausen2012a}.}
\label{spin_tr}
\end{figure}

Non-uniform magnetic fields are also at the root of a new type of spin transistor made of DMS QW, which has been demonstrated in Ref. \cite{Betthausen2012a}. The first spin transistor was proposed by Datta and Das \cite{Datta1990}. In this proposal, a spin polarized current is injected from the ferromagnetic source into a narrow channel of 2DEG where a gate-controlled SO field induces precession of the carrier’s spins. If the spins of carriers approaching ferromagnetic drain-analyzer after precession are aligned parallel to its magnetization, then the transistor is in the “on” state, if they are aligned antiparallel, the transistor is “off”. However, low efficiency of spin-injection and short spin lifetimes result in a quite small signal level, which limits practical realizations of such a transistor. In Ref. \cite{Betthausen2012a} a new type of spin transistor was proposed. The operation of such a spin transistor relies on diabatic Landau-Zener transitions that are controlled by a combination of the homogeneous, externally applied magnetic field and stray fields from a magnetized array of Dysprosium (Dy) stripes (see Fig.\ref{spin_tr}a).

In order to create stray fields, Dy stripes were magnetized at a high magnetic field. The stray fields form a modulated pattern, which slowly rotates the spin of the propagating electron. If the strength of the applied external magnetic field is lower that of the stray fields between the Dy stripes, then the electron spin slowly, adiabatically rotates in the superposition of external and stray fields and remains in the ground state throughout its motion along the channel, Fig.\ref{spin_tr}c. The spin transistor is in the “on” state. The situation is different if the external and stray fields cancel each other between Dy stripes. In this case, a degeneracy point occurs between the Dy stripes and the electron feels a discontinuity in the direction of the total field, Fig.\ref{spin_tr}d. Thus, in order to maintain the same spin direction, the electron would have to transfer to the upper Zeeman band. Since the potential energy rises above $E_F$, this leads to back-scattering and spin transport is blocked. The transistor is in the “off” state. Creation of a Dy stripes grating increases the probability of such a process giving rise to a noticeable MR effect. Indeed, for a period of 1~$\mu m$ between magnetized Dy stripes, a pronounced maximum of $\approx10\%$ in MR was observed at $\approx0.07~T$ (see Fig.\ref{spin_tr}b).

Worth mentioning are also theoretical papers proposing deposition of a type-II superconductor layer on top of the DMS QW containing 2DEG \cite{Berciu2005,Rappoport2006}. In a clean (i.e., without pin centers) superconductor placed in the magnetic field $\geq H_{c1}$ vortices form a triangular lattice which is a source of a periodic modulation of the magnetic field in the QW plane. This small modulation of the magnetic field creates spin and charge texture due to the giant spin-splitting energy. Such a periodic charge texture, being an image of the vortex lattice, forms a superlattice in the 2D plane, which leads to the phenomenon of Hofstadter butterfly \cite{Hofstadter1976} in a triangular lattice \cite{Hasegawa1990}. In a 1D system (e.g., a narrow superconductor stripe of the width $w\sim\lambda$, where $\lambda$ is the penetration depth), such periodic modulation would lead to Bloch oscillations, whose period will depend on the applied magnetic field. Though similar phenomena were studied in different systems \cite{Albrecht2001,Dean2013,Ponomarenko2013,Hunt2013}, the proposed platform may offer an interesting alternative for studying these effects since vortex lattices can be manipulated both with applied magnetic and electric fields. 

In Ref. \cite{Bruno2004} a 2-dimensional triangular lattice of ferromagnetic nanocylinders was proposed to induce the so called topological field in the DMS QW. An anomalous Hall contribution, called topological Hall effect, would arise due to electron motion in the topological stray field from magnetized nanocylinders. This topological field is zero on average but has some spatial variation which is solely determined by geometrical parameters of the nanocylinder array. Topological Hall effect, which arises in such a system, is expected to have a very distinct field dependence with several jumps of the Hall conductivity around field values which correspond to the change of topology of the stray field.

We also note that in the FQHE regime CFs are moving in an effective magnetic field $B^*=(1-2p\nu^*)B$ which depends on the local carrier density. Thus, any spatial variation of carrier density (e.g., created by non-uniformity of remote doping) will result in the inhomogeneous $B^*$ and hence modification of magnetotransport behavior, as observed in GaAs QWs \cite{Smet1998,Smet1999}. However, in the case of DMS QWs, there are no experimental or theoretical studies of the effect of giant spin splitting on CF magnetotransport in inhomogeneous $B^*$.

\subsection{DMS QWs under terahertz and microwave radiation}

\subsubsection{Radiation induced spin currents}

Many different methods have been proposed to generate spin currents, which are at the heart of spintronics, including the spin Hall effect (SHE). In the context of DMS QWs very interesting is the method of generating spin currents by terahertz (THz) \cite{Ganichev2006} or microwave \cite{Drexler2010,Olbrich2012} radiation. This approach is based on the spin dependent scattering of carriers by static defects or phonons. In gyrotropic media (such as e.g., GaAs or CdTe QWs) an asymmetric, spin-dependent term due to SO interaction exists in the scattering probability. This additional term is linear in respect to the wavevector {\bf k} and the Pauli matrices {\bf{$\sigma$}}. Thus, both at the stage of Drude absorption of radiation by carriers followed by scattering (polarization dependent excitation mechanism), and at the stage of energy relaxation process of heated carriers (relaxation mechanism) an imbalance in the distribution of photoexcited spin-up and spin-down carriers between positive and negative {\bf k} is created. Each of the equally occupied spin subbands creates an equal electric current, but with the direction that depends on the spin orientation. Therefore, electric currents from spin-up and spin-down subbands cancel each other leaving a pure spin current.

These spin currents can be converted into spin-polarized electrical currents by applying an external magnetic field, which polarizes the carriers and reduces the compensation of electrical currents from opposite spin subbands. Magnetic-field-induced electrical currents are determined by spin imbalance in the system and, consequently, by the spin-splitting energy \cite{Ganichev2006,Ganichev2009}: ${\bf j}_x=-e\frac{E_z+E_{exch}}{k_BT}{\bf J}_s$, where ${\bf J}_s$ is the spin current and ${\bf j}_x$ is the converted electrical current. Indeed, using QWs with higher $g^*$ (e.g., narrow-gap InAs QW), and especially DMS-based QWs (e.g. (Cd,Mn)Te chalcogenide QW) with giant effective $g$-factor $g_{\text{eff}}$, results in very strong amplification of the spin to electrical current conversion \cite{Ganichev2006,Ganichev2009,Drexler2010,Olbrich2012}. In (Cd,Mn)Te QWs this amplification occurs for two reasons \cite{Olbrich2012}. The first is the giant spin splitting, which determines redistribution between spin subbands, and the second reason is the spin-dependent scattering of carriers on $\text{Mn}^{2+}$ spins which are polarized in an applied magnetic field. Exchange contribution to the spin splitting is also responsible for the particular temperature dependence of the converted electrical current, which reverses its sign upon cooling to helium temperatures. That is because exchange spin-splitting $E_{exch}$ depends on the Brillouin function $\mathfrak{B}_s$ (eq. (\ref{eqLL_exch})), and at elevated temperatures, when $\mathfrak{B}_s$ is negligible, the spin current is determined solely by the intrinsic Zeeman term $E_z$. At low temperatures, however, the exchange term takes over reversing the sign of the spin current and strongly increases its amplitude.

\subsubsection{Magnetic quantum ratchet effects}

Now we will turn to THz radiation induced photocurrents (ratchet currents) in the 2DEG system subjected to spatially periodic, non-centrosymmetric, lateral potentials \cite{Ivchenko2011}. Such potential can be achieved, e.g., by a metallic, asymmetric lateral dual-grating gate superlattice produced at the surface of the 2DEG structure. This grating screens the 2DEG underneath the metallic fingers from the THz radiation, while the unprotected 2DEG is heated. Thus, a metal grating leads to inhomogeneous heating of 2DEG inducing modulation of the local electron temperature. Because of the temperature gradient, electrons diffuse from the hot to cold regions forming a non-equilibrium density profile. The ratchet current at zero magnetic field can be represented as a drift current of the electrons in the electric field of the spatially modulated electrostatic potential. This mechanism of creating polarization independent ratchet current, called {\it thermoratchet} (or Seebeck ratchet) \cite{Blanter1998}, dominates at zero magnetic field and is also most relevant to the magnetic quantum ratchet current observed in DMS QWs \cite{Faltermeier2017}.

In the absence of magnetic field the position dependent relative correction to the concentration, which defines ratchet current, is much smaller than the temperature correction, which results in a rather weak ratchet current. However, the situation changes when the magnetic field is applied \cite{Budkin2016,Faltermeier2017}. The magnetic field induces ratchet currents in directions both perpendicular and parallel to the stripes of the metal grating. In the presence of quantizing magnetic fields ratchet current exhibit $1/B$-periodic oscillations with the same period as SdH oscillations and has strongly enhanced magnitude as compared to ratchet current at zero field. This follows from the fact that now the photocurrent arises from heating induced correction to conductivity rather than from correction to electron density \cite{Faltermeier2017}. Near the Dingle temperature variations of the electron temperature result in an exponential increase of conductivity, thus making ratchet current much stronger. Oscillations of the ratchet currents in magnetic and non-magnetic QWs are different, because of the alteration of the LL diagram caused by the magnetic impurities (Fig.\ref{fLLfan}). Strong temperature-dependent modulation of $1/B$-periodic oscillations of ratchet currents in (Cd,Mn)Te QWs were reported in Refs. \cite{Faltermeier2017,Faltermeier2018}. Inclusion of the exchange term $E_{exch}$ in the spin-splitting energy for magnetic QWs allowed also to explain such experimental findings as the beating pattern of oscillations and weakening of the high field ratchet current at low temperatures. None of these three characteristics of magnetic quantum ratchet have been observed in non-magnetic CdTe QWs.

\section{Novel topological phases in chalcogenide multilayers}

In recent years there has been a burst of interest in new phases of quantum matter, which are characterized by a topological invariant of their ground state \cite{Hasan2010,Qi2011}. These phases are topologically protected, which means that they are protected against any perturbation by a certain symmetry. One of the examples of such materials is the topological insulator (TI), which has non-trivial boundary states that are protected against back-scattering by time-reversal symmetry \cite{Qi2011}. Surface and edge states of topological insulators can find application in spintronics because of the spin-momentum locking \cite{Hasan2010,Ando2013}. Another gapped topological state, namely the topological superconductor, can be used in the field of quantum computation \cite{Nayak2008,Mong2014}. Boundary states of topological superconductors obey non-Abelian statistics and are known in the literature as Majorana fermions \cite{Kitaev2001,Alicea2012,Beenakker2013,Alicea2015}. More information about topological insulators (TIs) can be found in other chapters of this book. In the current chapter we will focus on different proposals \cite{Fatin2016,Simion2018} on creating Majorana fermions in DMS QWs and briefly mention 2D topological phases which were proposed to exist \cite{Bernevig2006a,Liu2015,Safaei2015} in II-VI and IV-VI semiconductor heterostructures and the ones, which were experimentally found \cite{Koenig2007,Budewitz2017} in HgTe QWs.

\subsection{Domain walls and non-Abelian excitations}

\begin{figure}[t]
\centering\includegraphics[width=0.75\columnwidth]{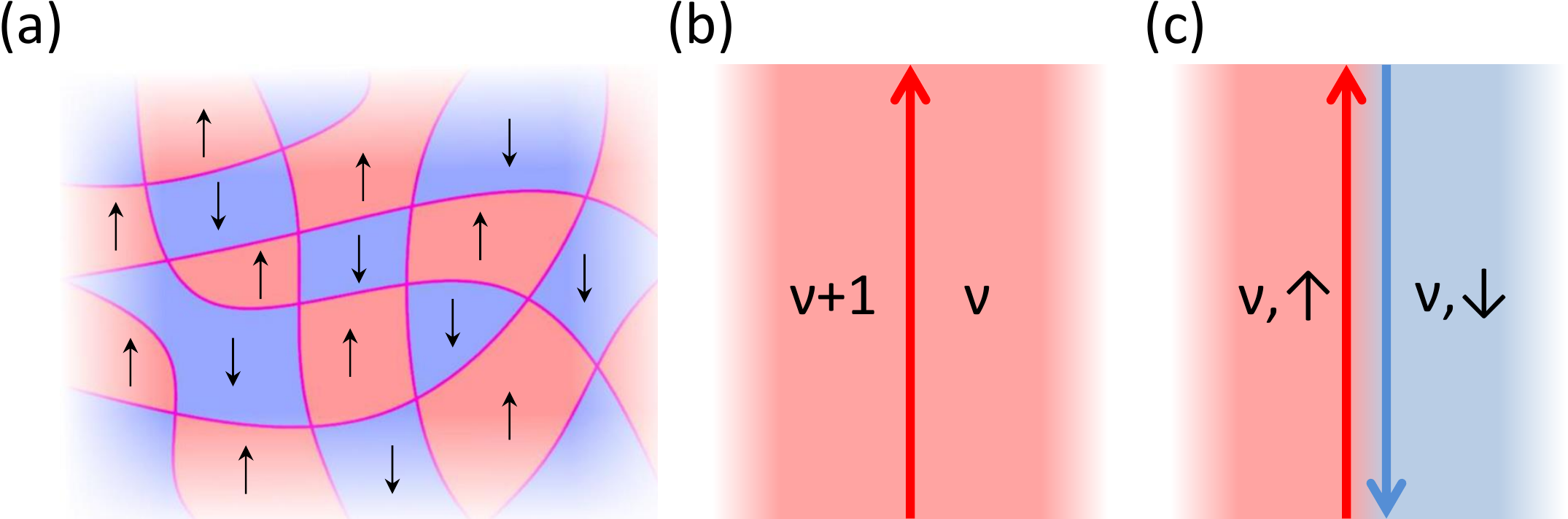}
\vspace{0in}
\caption{(a) Domain structure during QHF transition. Arrows and color denote spin polarization of the top filled LL. (b) {\it Chiral} domain boundary between QHE states with different $\nu$. (c) {\it Helical} DW between QHE states with different spin polarization but the same $\nu$. It consists of two counter-propagating edge channels with opposite spins.}
\label{fDW}
\end{figure}

As we discussed in the previous section, QHF transition is accompanied with an additional maximum in $R_{xx}(B)$. According to the QHE picture, it means that 2DEG goes through a conductive state. It was argued that during such transition a network of conductive domain walls (DWs) is formed and conduction occurs through these conductive channels \cite{Falko1999,Mitra2001,Jungwirth2001,Chalker2002}. These DWs separate domains of opposite spin polarization of top filled LL (Fig. \ref{fDW}a), and they are not chiral anymore since they are separating QHE states with the same filling factors --- in the simplified picture, they consist of two counter-propagating chiral channels, which correspond to LL with opposite spins (Fig. \ref{fDW}c). In such a viewpoint, these DWs are helical \cite{Kazakov2017,Simion2018}, similar to edge states of TIs. There was speculation about possible transport properties of a single DW in the theoretical works at the beginning of 2000's \cite{Falko1999,Mitra2001,Jungwirth2001,Chalker2002,Brey2002}, but no experimental study of a single DW was done at that time.

The study of an individual DW is a challenge since the conventional way to induce the QHF transition is either by tilting $B$ \cite{DePoortere2000,Teran2010} or by changing the magnetic dopant concentration \cite{Wojtowicz2000}. Both approaches drive the QHF transition in the entire 2DEG, controlling thus the {\it global} spin polarization of the QHE system. {\it Local} control of spin polarization becomes possible in DMS QWs with magnetic ions introduced asymmetrically to the QW region along the growth direction. It was shown \cite{Kazakov2016} that in such a structure the electric field created by an electrostatic gate changes the $E_{exch}$ contribution to the spin-splitting energy (eq. (\ref{eqLL_exch})). This becomes visible when considering $E_{exch}$ in the mean-field approximation of the exchange Hamiltonian: $J_{sd}\sum_{{\bf R}_i}\delta({\bf r}-{\bf R}_i){\bf S}_i\cdot {\bf \sigma}\propto \left[\int_{[\text{Mn}]}|\varphi(z)|^2dz\right] \langle{\bf S}\rangle$, where interaction of an electron at a position ${\bf r}$ with a large number of Mn ions at positions ${{\bf R}_i}$ is approximated as an overlap of the electron probability density $|\varphi(z)|^2$ with a uniform Mn background within $z\in[\text{Mn}]$ and an average magnetization $\langle{\bf S}\rangle=\langle S_z\rangle=S\mathfrak{B}_s(B,T)$. Within this approach it is easy to see that in QWs with homogeneous Mn distribution the overlap of the electron wave function with $\text{Mn}^{2+}$ ions does not change, even when the wave function is shifted due to the electric field created by voltage applied to the gate, provided that the potential barrier is high enough, so that the wave function does not penetrate the barrier material. Thus exchange splitting remains unchanged \cite{Jaroszynski2002}. On the contrary in QWs with Mn introduced only into some specific region of QWs an application of gate voltage shift the electrons with respect to this Mn containing region, changes the overlap integral and hence changes exchange contribution to the effective $g$-factor $g_{\text{eff}}$ of 2DEG (either it increasing or decreasing).

\begin{figure}[t]
\centering\includegraphics[width=0.75\columnwidth]{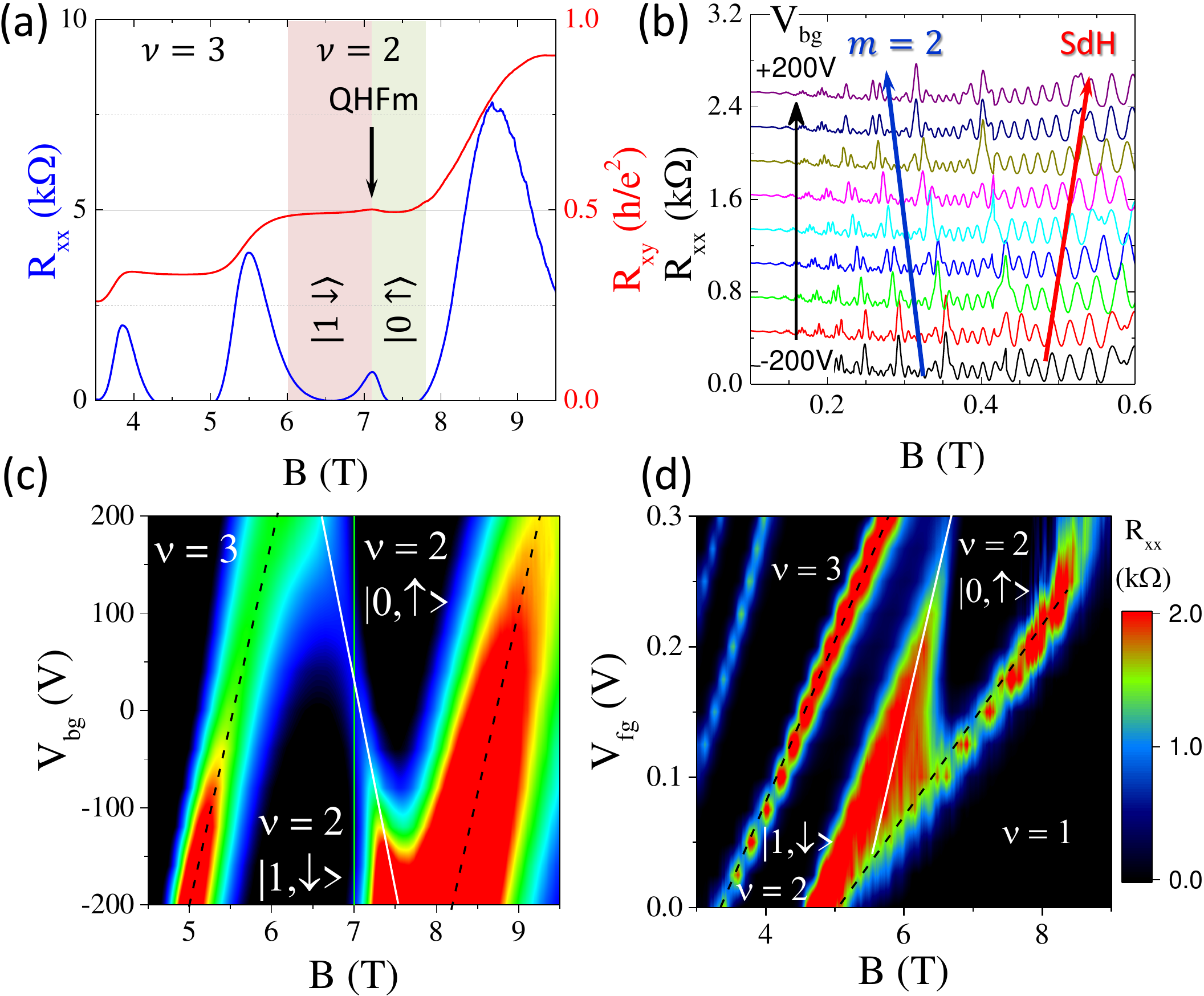}
\vspace{0in}
\caption{(a) The QHF transition is clearly seen in the longitudinal resistance ($R_{xx}$) at $\nu=2$ QHE state (seen from Hall curve ($R_{xy}$)), measured in the (Cd,Mn)Te QW with $x_{\text{eff}}\sim1.7\%$ for zero gate voltages. The QHF transition separates $\ket{1\uparrow}$ and $\ket{0\downarrow}$ states. (b) SdH beating pattern clearly changes its shape with variation of the back gate voltage from -200~V to +200~V in the (Cd,Mn)Te QW with $x_{\text{eff}}\sim0.2\%$. Nodes are shifting to lower fields (blue arrow) and SdH extrema are shifting to higher fields (red arrow) with increasing the back gate voltage. (c,d) Evolution of the QHF peak shown in (a) under application of back (front) gate voltage with fixed front (back) gate voltage, respectively. The position of the QHF transition is highlighted by the white line. In (c) the green line marks the value of $B$ for which polarization of the top LL can be switched between $\uparrow$ and $\downarrow$ by gate voltage. Reprinted with permission from Ref. \cite{Kazakov2016}.}
\label{fexch_control}
\end{figure}

Indeed, it was demonstrated \cite{Kazakov2016} that applying a gate voltage to asymmetrically Mn doped QWs changes $g_{\text{eff}}$ of a 2DEG. This voltage control of $g_{\text{eff}}$ was revealed using two manifestations of magnetic interaction in the QWs discussed in the previous section, i.e., low-field SdH beating pattern (for low Mn content, Fig.\ref{fexch_control}b) and high field additional QHF resistance cusp (for larger Mn content, Fig.\ref{fexch_control}a). Thus, electrostatic gating can be used to define an isolated DW at the gate edge, which separates LL of opposite spin polarization.

The first study of DW transport properties was reported in Ref. \cite{Kazakov2017} where a (Cd,Mn)Te QW with Mn introduced asymmetrically was used to form an isolated DW. In this experiment, samples were patterned into alternating gated and ungated Hall bar sections. These sections were separated by narrow constrictions along which gate boundaries were aligned. The lithographical length of the DW formed along the gate boundary in the constriction varied with constriction size from $\sim8~\mu m$ to $\sim0.8~\mu m$. This experiment employed QHF transition at the $\nu=2$ QHE state, i.e., LL crossing between $\ket{0\uparrow}$ and $\ket{1\downarrow}$ states (Fig.\ref{fLLfan}b). Transport properties of the single DW were probed indirectly by measuring a longitudinal resistance $R_{\text{DW}}$ across the constriction in the presence of a phase boundary.

\begin{figure}[t]
\centering\includegraphics[width=0.75\columnwidth]{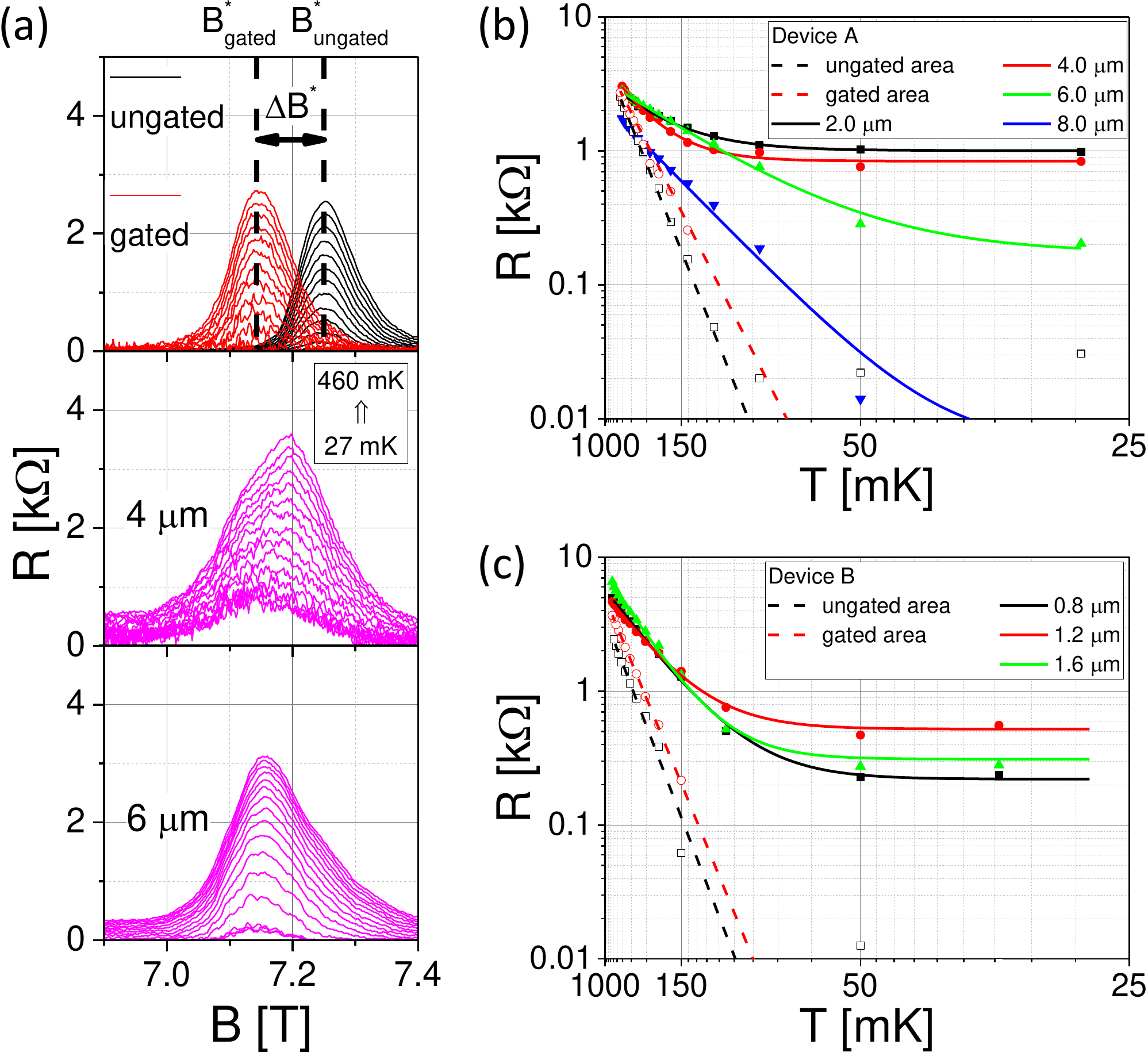}
\vspace{0in}
\caption{(a) In the devices designed for studying DW transport in the QHE regime $R_{xx}$ was measured both in the large ungated and gated areas of the Hall bar (upper panel), at the same time $R_{\text{DW}}$ was measured across a narrow constriction in the presence of a DW. In the middle panel, the saturating behavior of $R_{DW}$ in the 4~$\mu$m constriction and in the lowest panel the vanishing of $R_{DW}$ in the 6~$\mu$m constriction upon lowering temperature are presented. (b,c) Arrhenius plots for the temperature dependencies $R_{\text{DW}}(T)$ in two different devices for various size constrictions. Solid lines are fits to $R=R_0+A\cdot e^{-\Delta/2kT}$, dashed lines are fits to thermally activated conduction in large areas. Reprinted with permission from Ref. \cite{Kazakov2017}.}
\label{fDW_R}
\end{figure}

The wide Hall bar sections enable independent measurements of resistance spikes associated with QHF transitions in ungated and gated regions (upper panels in Fig.\ref{fDW_R}a). At low temperatures, QHF resistance spikes vanish in these regions. For large size constrictions (thus long DWs) the voltage drop also vanishes at low temperatures (the lowest panel in Fig.\ref{fDW_R}a). In contrast, for narrow constrictions (thus short DWs with lengths of $<4~\mu m$), the voltage drop across the constriction saturates at low temperatures (middle panel in Fig.\ref{fDW_R}a). This indicates that a conductive channel is formed along the gate boundary. However, the low-temperature saturation value of $R_{xx}$ does not depend on the constriction size (Fig.\ref{fDW_R}b,c). It was also shown that these channels exhibit the necessary for {\it helical} channels symmetry with respect to magnetic field reversal. Under reversed magnetic fields the resistance of these DWs remains the same \cite{Kazakov2017}, while {\it chiral} channels formed at the boundary of QHE states show quantized resistance in one field direction and zero resistance in the opposite direction \cite{Haug1993}.

\begin{figure}[t]
\centering\includegraphics[width=0.75\columnwidth]{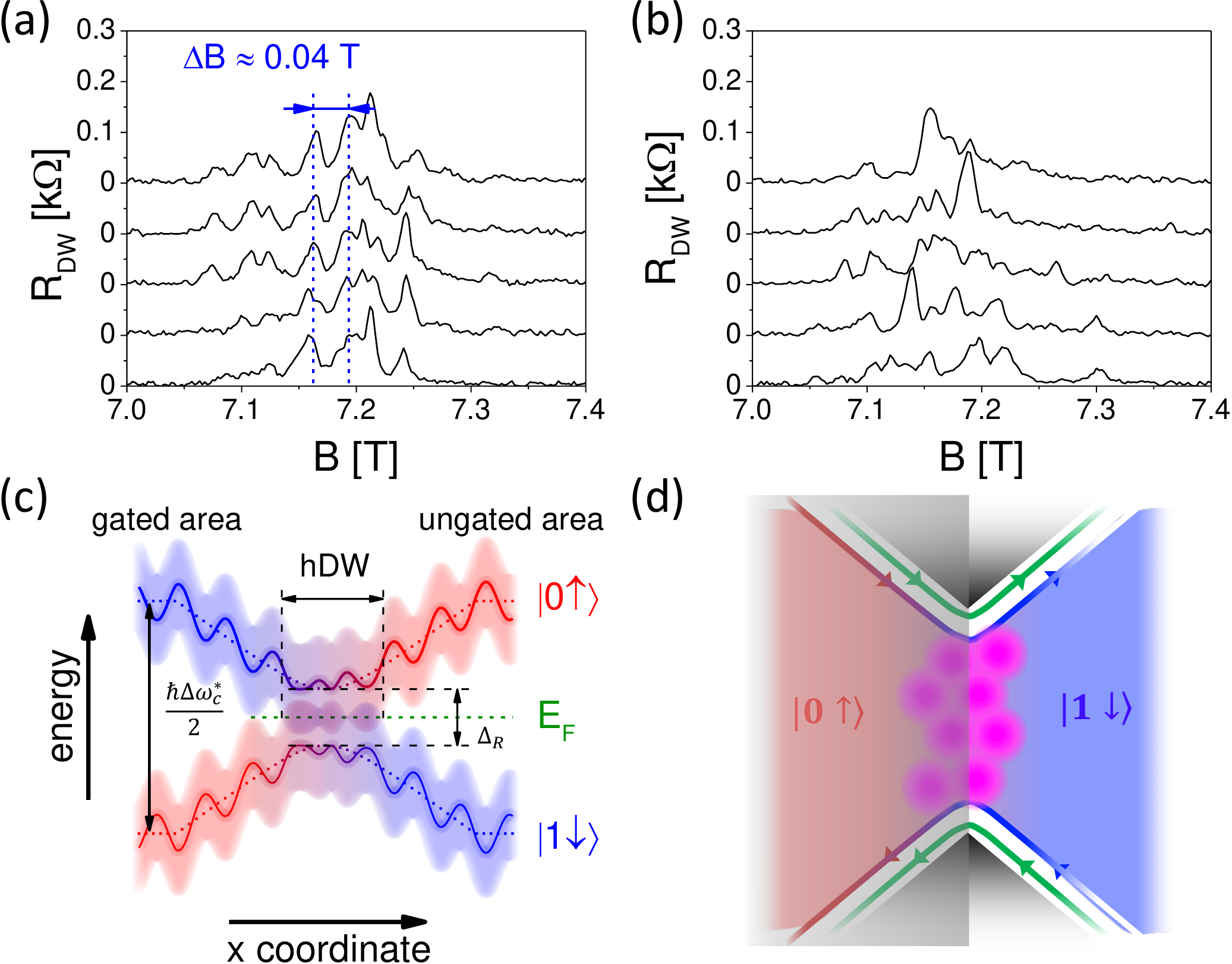}
\vspace{0in}
\caption{(a,b) $R_{\text{DW}}$ measured for a 2~$\mu m$ constriction exhibits mesoscopic fluctuations. The resistance pattern remains similar if the magnetic field is repeatably swept in a narrow range within $\nu=2$ (a) and changes considerably if $B$ was changed is swept in a wider range, $5-10$~T (b). (c) Energy diagram of a DW formed at the gate boundary. Wiggling lines indicate schematically the role of disorder and shaded areas are localized states in the tails of LLs. At low temperatures, conduction occurs via localized states in the gap. (d) Schematic representation of the conducting channel formed by coupled $\nu=2$ edge states. Electron tunneling via in-gap states (magenta) provides several interfering trajectories resulting in observed mesoscopic fluctuations of $R_{\text{DW}}$ in (a) and (b). Dark shaded area represents a gate. Adapted with permission from Ref. \cite{Kazakov2017}.}
\label{fDW_fluct}
\end{figure}

Simple model calculations following Landauer-B{\"{u}}ttiker formalism \cite{Buettiker1988} showed, that these channels have relatively high resistance at low temperatures --- $>h/e^2$. Nevertheless, electronic transport through short DWs shows mesoscopic fluctuations (Fig.\ref{fDW_fluct}a,b), which are clearly seen at low temperatures. Analysis of these fluctuations allowed extracting phase coherence length, which value was comparable to the length of the DW. It was found that the fluctuation pattern changes drastically if the magnetic field is ramped outside the $\nu=2$ QHE state (Fig.\ref{fDW_fluct}b). It was concluded that dynamic fluctuations rather than static impurities define an actual conduction path in the DW. Activation type behavior of electronic transport through the DWs suggests that its energy spectrum is also gapped, as that of 2DEG LLs. Together with the mesoscopic fluctuations and the suppression of the conduction in long DWs at low temperatures, it suggests the following mechanism of conduction in the DW.

At low temperatures electron states inside the energy gap become localized, thus suppressing transport through long DWs ($> 6~\mu m$). However, if the length of DW is less than the localization length, transport through localized in-gap states becomes possible even at low temperatures. This scenario is visualized in Fig.\ref{fDW_fluct}c, where a single DW is formed by the localized states in the SO-induced anticrossing gap between broadened LLs. These localized states provide a conduction path which defines observed mesoscopic fluctuations (Fig.\ref{fDW_fluct}d).

An isolated DW offers the potential opportunity to create and manipulate non-Abelian excitations, namely Majorana quasi-particles \cite{Simion2018} which are important for topological quantum computation \cite{Nayak2008,Mong2014}. In Ref. \cite{Simion2018} a hybrid structure was considered, which consisted of {\it s}-superconductor and QHF DW at the filling factor $\nu=2$. The magnetic field, as well as the global gate voltage, were set to bring the system close to the QHF transition. $E_F$ was in the middle of the SO induced gap so the 2DEG conductivity, except DWs, was exponentially suppressed at low temperatures.

As it was discussed previously, in-gap impurity states are responsible for conductivity along short enough DW. Each impurity generates two energy levels for every LL, so in the vicinity of QHF transition at $\nu=2$, there are four energy levels bound to the impurity, which are generally non-degenerate unless cyclotron splitting is compensated by the sum $E_z+E_{exch}$. In the case of such compensation, the impurity-induced states form pairs, each doubly degenerate. Along a DW these impurities form a spatial chain of in-gap states, which is responsible for the conductivity at low temperatures. Together with superconducting pairing, it results in topological superconductivity along the DW. The resulting system resembles generalized Kitaev chain \cite{Sau2012} that supports two Majorana localized modes at its ends if
\begin{equation}
\lvert\mu_k-L_k\rvert<max(t_{k,k+1},\Delta_{k,k+1}),
\label{Majorana_cond}
\end{equation}
where $k$ denotes the spatial index of the impurity along DW; $\mu_k$ is the chemical potential at the impurity; $L_k$ is the angular momentum dependent on the gate-induced potential gradient (the latter parameter is also responsible for removing fermion doubling in the DW); $t_{k,k+1}$ is the effective tunneling amplitude between adjacent impurity-induced states; $\Delta_{k,k+1}$ is the induced superconducting pairing. Thus, topological superconductivity can be controlled exclusively by the gate voltage gradient, provided that DW is proximatized by superconducting contacts. With an appropriate multiple gate structure, Majorana modes can be manipulated, fused and braided \cite{Simion2018}.

In the end, it should be noted that similar physics is expected in case of FQHE. The difference is that in the FQHE regime and with induced superconductivity DWs in the vicinity of a spin transition would lead to the formation of higher order non-Abelian excitations --- parafermions \cite{Simion2018,Wu2018a,Liang2019}.

\subsection{Wireless Majorana bound states}

In Ref. \cite{Fatin2016} another interesting proposal for creating and braiding non-Abelian excitations was presented. The proposal was based on using a semiconductor with a large $g^*$, such as e.g., (Cd,Mn)Te QW, and taking advantage of its giant spin splitting. In the modeled system a superconducting gap in the DMS QW was induced through an s-wave superconductor located underneath the 2DEG. A two-dimensional array of magnetic tunnel junctions (MTJs) was placed on the top of the QW.

MTJs can be in two states: either ON (parallel configuration) or OFF (antiparallel configuration). Topologically nontrivial regions are created by inducing a spatial gradient of the perpendicular magnetic field. Thus, magnetic texture produced by the switchable magnetization of MTJ provides a way to control topological phase transitions for confining and braiding Majorana bound states (MBSs). A chain of MTJs in the ON state with the alternating direction of vertical magnetization creates an effective 1D wire which supports MBSs on its ends. By switching the magnetization of a single MTJ from OFF to the proper ON state and vice versa one can move, braid, and fuse MBSs. By measuring the change of a charge (probed either by single-electron transistor spectroscopy or by STM) in a single quantum dot under an initialized MTJ non-Abelian statistics can be probed. An interesting feature in this proposal is the absence of rigid 1D channels: all manipulations with MBSs are performed purely in the 2D plane.

\subsection{Quantum spin Hall effect in HgTe QWs} \label{sQSHE_HgTe}

HgTe QWs have recently attracted a lot of attention, since they were first proposed \cite{Bernevig2006} and then confirmed experimentally \cite{Koenig2007} to be 2D topological insulators. A topological insulator (TI) is a novel state of matter which is insulating in the bulk but has gapless states at its boundaries, i.e., edge states in 2D and surface states in 3D TIs, respectively.

\begin{figure}[t]
\centering\includegraphics[width=0.75\columnwidth]{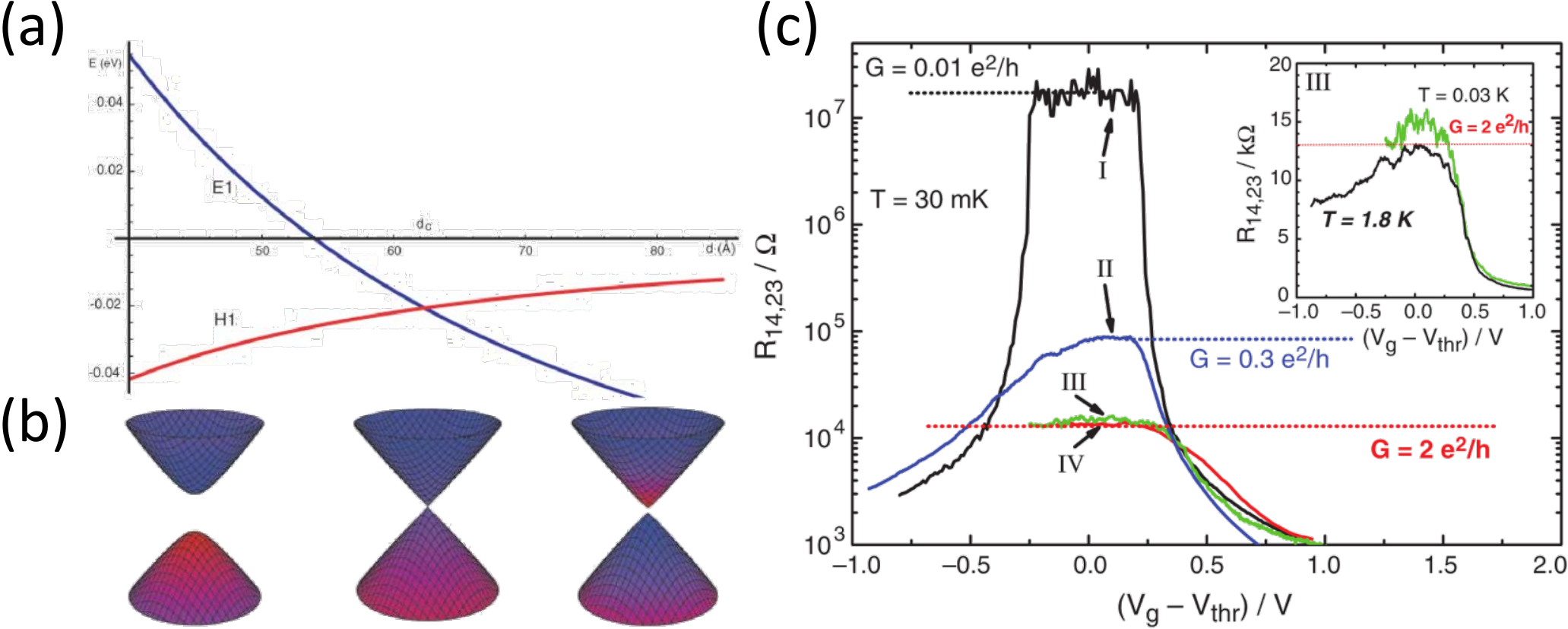}
\vspace{0in}
\caption{(a) Dependence of E1 and H1 energy bands on HgTe QW thickness $d$. (b) Energy dispersion of E1 and H1 bands at thicknesses corresponding to the trivial regime (left), band crossing at critical QW thickness $d_c$ (center), and to the inverted regime (right). The color scheme indicates the type of symmetry and is the same as applied in (a). Reprinted with permission from Ref. \cite{Bernevig2006}. (c) Experimental evidence for QHSE in HgTe/CdTe QW. For Hall bars of tens of microns size (devices I and II) an insulating behavior is observed (two terminal conductance $G<1~e^2/h$) while for Hall bars of one micron and less (devices III and IV) conductance quantization $G=2~e^2/h$ is observed. Reprinted with permission from Ref. \cite{Koenig2007}.}
\label{fQSHE}
\end{figure}

The central role in the theoretical proposal for realizing 2D TI and its hallmark, quantum spin Hall effect (QSHE) \cite{Bernevig2006} is the band inversion. It was already known \cite{Harman1964,Piotrzkowski1965} that the barrier material CdTe has normal band alignment with the s-type $\Gamma_6$ band lying above the p-type $\Gamma_8$ band, while in HgTe the band alignment is inverted. The $\Gamma_6$ band lies below the $\Gamma_8$ band. In HgTe/CdTe QWs these bands form E1, H1, and L1 subbands (the latter is separated from the first two and can be disregarded). In a HgTe/(Hg,Cd)Te QW with thickness $d$ lower than the critical thickness $d_c\simeq6.3$~nm "normal" band arrangement occurs (E1 lies above H1), as in CdTe, but when $d>d_c$, the band structure is inverted ($E1$ lies below $H1$)\cite{Bernevig2006}, see Fig.\ref{fQSHE}a-b.

In the 'inverted' regime (QSHE state) "helical" edge states appear at the sample edges. Their main property is that they are "spin filtered", i.e., the up spin propagates in one direction, while the down spin propagates in another. A quantized value of two-terminal conductance $G=2e^2/h$ was noted as an experimental consequence of QSHE \cite{Kane2005}. Indeed, a value of two-terminal conductance close to the predicted quantized value was observed experimentally \cite{Koenig2007} in the $7.3$~nm thick $\text{HgTe}/\text{Hg}_{0.3}\text{Cd}_{0.7}\text{Te}$ QW at zero magnetic field (Fig.\ref{fQSHE}c), however, only over distances shorter than $1~\mu m$. Longer distances lead to a vanishing conductivity of edge channels \cite{Koenig2007,Grabecki2013}. Meanwhile, theoretical predictions \cite{Kane2005a} state that helical edge states are topologically protected against back-scattering unless the time-reversal symmetry is broken. Different mechanisms \cite{Hsu2017,Groenendijk2018,Wang2017,Vaeyrynen2016,Vaeyrynen2014,Stroem2010,Tanaka2011,Budich2012,Vaeyrynen2013} of inelastic back-scattering were proposed, which limit the length over which quantized transport can be observed. However, the mechanism responsible for inelastic backscattering in helical channels has not been yet determined experimentally.

One of the exciting proposal to implement TI helical edge states in future devices is to use them as a platform for the realization of topological superconductivity \cite{Fu2009}. That's why the investigation of the interplay between 2D TI phases and superconductivity attract a lot of attention in recent years \cite{Hart2014,Hart2016,Ren2019}. HgTe QWs are one of the main focuses of current research in this field. It was shown by studying the Fraunhofer interference that Josephson supercurrent indeed can be induced to QSHE edge states \cite{Hart2014}. Later study revealed details of the electron pairing, which occurs in the materials with non-trivial spin structure proximatized by a superconductor \cite{Hart2016}. Recently it was experimentally shown that induced superconductivity has indeed a different nature in trivial and topological regimes, which was controlled by applying in-plane magnetic field \cite{Ren2019}. Thus, HgTe QWs are a promising platform in building scalable networks of Majorana devices for fault-tolerant quantum computation.

\subsection{Quantum anomalous Hall effect in HgTe QWs} \label{sQAHE_HgTe}

It was predicted \cite{Liu2008b} that magnetic doping may break time-reversal symmetry and leave only one spin-polarized edge state out of the two forming QSHE helical edge state. This can be understood as follows. Magnetic doping introduces a spin-splitting term to the QSHE Hamiltonian and if one of the spin blocks moves to the trivial regime while the other one remains in the 'inverted' regime, then the system will be in so-called QAHE state. This imposes certain constriction on the spin-splitting gaps $2G_E$ and $2G_H$ of E1 and H1 states of HgTe QW respectively, i.e. $G_EG_H<0$. This condition simply means that $g^*$ of E1 and H1 subbands must have opposite signs, which fortunately is the case for HgTe QWs \cite{Koenig2008a}. Analysis of the QAHE parameter space for Mn-doped HgTe QW \cite{Liu2008b} showed that it can be in the QAHE state as long as the magnetization is large enough and perpendicular to the QW plane.

The main problem for QAHE realization in (Hg,Mn)Te/CdTe QWs is the fact that the exchange field does not drive the magnetic system into the ferromagnetic state since QWs are paramagnetic up to Mn content of $\approx15\%$ \cite{Brandt1984}. Thus, a small external field ($\approx70~mT$) is still required to drive the system to the $\nu=-1$ QAHE state \cite{Budewitz2017}. However, any small magnetic field has also an orbital effect and such observation alone cannot unambiguously prove the existence of the QAHE state. Nevertheless, experiments in tilted magnetic fields showed different evolution of the $\nu=-1$ state under an in-plane field in nonmagnetic HgTe and magnetically doped (Hg,Mn)Te QWs. This observation indicates that these states have a different origin.

\subsection{Topological phases in IV-VI materials} \label{sTCI}

In recent years there is a renewed interest in IV-VI semiconductors containing Tin (e.g., (Pb,Sn)Se). It was predicted \cite{Hsieh2012} and then confirmed by ARPES measurements \cite{Dziawa2012,Tanaka2012} that they belong to a new class of topological insulators --- topological crystalline insulators (TCIs), whose surface states are topologically protected by crystalline symmetries \cite{Fu2011,Hsieh2012}. Calculations show that while $\text{Pb}_{1-x}\text{Sn}_{x}\text{Te}$ and $\text{Pb}_{1-x}\text{Sn}_{x}\text{Se}$ are topologically trivial for $x_{\text{Sn}}=0$ \cite{Hsieh2012}, band inversion at low temperatures occurs with increasing Sn content at critical values of $x^c_{\text{Sn}}=0.35$ \cite{Dimmock1966} for tellurides and $x^c_{\text{Sn}}=0.15$ for selenides \cite{Calawa1969}. It was shown theoretically that a QW structure made of TCI material will also host helical edge states \cite{Liu2015,Safaei2015}, and in the case of a ferromagnetic order --- the QAHE phase with a higher Chern number \cite{Fang2014}. Topologically trivial nature of PbTe/PbEuTe QWs was established in Ref. \cite{Kolwas2013} by careful analysis of non-local measurements interpreted in terms of parallel conductance through PbEuTe barriers. However, studies of $\text{Pb}_{1-x}\text{Sn}_{x}\text{Te}$ or $\text{Pb}_{1-x}\text{Sn}_{x}\text{Se}$ QWs with the composition in the topological regime still await realization.

\section{Summary and perspectives}

Despite being studied for almost 30 years, 2D systems based on chalcogenide multilayers still attract a lot of attention. Almost all recent developments concerning 2DEG reviewed here are based on the fact that magnetic ions can be easily incorporated into the chalcogenide QW structure, thus giving rise to giant spin splitting due to exchange interaction between carriers and magnetic ions. Moreover, with the use of MBE growth technique, spin splitting (or in other words $g_{\text{eff}}$) can be engineered not only by changing concentration of magnetic component in disorder alloys, but also by placing magnetic ions in strictly predefined positions, so as to produce digital alloys, graded potential QWs \cite{Wojtowicz2000} or magnetically asymmetric QWs \cite{Kazakov2016}. In the latter case, spin splitting can be externally controlled not only by magnetic field and temperature but also by the electric field from the voltage applied to electric gates in nanodevices \cite{Kazakov2017}. Very important is also the fact that incorporation of magnetic ions into chalcogenides, at already useful level of concentration of localized spins, does not lead to any strong reduction of 2DEG mobility \cite{Betthausen2014,Kunc2015}. Finally, the concentration of 2DEG can be varied by donor doping independently of magnetic doping. All these facts lay at the basis of many proposals to implement DMS QWs in future devices and realization of novel topological phases. Some of these proposals are still awaiting realization.

The very important direction is studying superconducting proximity effect in 2DEG in chalcogenide QWs. Several platforms for realization of a topological qubit, based on materials reviewed in the current chapter, have been proposed in recent years. The very interesting direction in this area is utilizing DWs which emerge during QHF transitions to create and manipulate non-Abelian excitations \cite{Simion2018}. This particular goal would also require inducing superconductivity in 2DEG through the proximity effect. Superconductivity induced in a semiconductor has attracted a lot of attention nowadays because of potential application for topological quantum computation \cite{Nayak2008,Mong2014}. Also, there are only a few studies of the interplay between a superconductor and QHE states \cite{Wan2015,Amet2016,Lee2017}. Investigation of such interplay would be particularly interesting in the case of DMS QWs with a giant spin splitting. Quite interesting would be a practical realization of topological phases in a non-uniform magnetic field created by a periodic lattice of vortices in a superconductor proposed already many years ago \cite{Berciu2005}. The increasing quality of grown structures would allow observing more fragile FQHE states, which versatile physics can also bring unexpected results in the case of 2DEGs with giant spin splitting. Moreover, the realization of 2D topological phases in QWs of IV-VI semiconductors will introduce another member to the small group of 2D TI materials \cite{Koenig2007,Knez2011,Wu2018}. There is also hope, that further improvement in the quality of heterostructures, which is to be achieved by progress in the MBE growth technology, will help to develop new ideas for interesting physical studies and for novel devices based on 2DEGs in chalcogenide multilayers.

\section*{Acknowledgement}

This work was partially supported by the Foundation for Polish Science through the IRA Programme co-financed by EU within SG OP and by the National Science Centre (Poland) through Grant No. DEC- 2012/06/A/ST3/00247.


\end{document}